\documentclass[11pt,tightenlines,eqsecnum,floats,aps,amsmath,amssymb,nofootinbib,prd,shownopacs,floatfix]{revtex4}
\usepackage[english]{babel}
\usepackage[utf8x]{inputenc}
\usepackage{amsmath,amsfonts,amssymb, extarrows,mathrsfs}
\usepackage{xcolor}
\usepackage{graphicx}
\usepackage{empheq}
\usepackage{hyperref}
\usepackage{bbm}
\usepackage{pdfpages}
\usepackage{makecell}
\usepackage{nicefrac}

\newcommand{\be}{\begin{equation}}
\newcommand{\ee}{\end{equation}}
\newcommand{\ba}{\begin{eqnarray}}
\newcommand{\ea}{\end{eqnarray}}

\DeclareMathAlphabet{\mathpzc}{OT1}{pzc}{m}{it}
\newcommand*\bigcdot{\mathpalette\bigcdot@{.5}}
\newtheorem{lemma}{Lemma}

\renewcommand{\d}{\mathrm{d}}

\newcommand{\tdif}{\ensuremath{\frac{\d}{\d t}}} 
\newcommand{\Ht}{\tilde{\mathcal{H}}} 
\newcommand{\Pt}{\tilde{P}} 
\newcommand{\Nb}{\bar{N}} 
\newcommand{\dT}{\delta\tilde{T}^{(\mathrm{gi})}} 
\newcommand{\lphi}{\lambda_\varphi} 
\newcommand{\dl}{\delta\lambda} 
\newcommand{\dlh}{\delta\hat{\lambda}} 
\newcommand{\dphi}{\delta\varphi} 
\newcommand{\dpiphi}{\delta\pi_\varphi} 
\newcommand{\piphibar}{\bar{\pi}_\varphi} 
\newcommand{\piv}{\pi_v} 
\newcommand{\Obs}[1]{\mathcal{O}_{#1,T}^{(1)}} 
\newcommand{\dotObs}[1]{\dot{\mathcal{O}}_{#1,T}^{(1)}} 
\newcommand{\appref}[1]{\ensuremath{({\rm C}#1)_{\text{\scriptsize\cite{Giesel:2017roz}}}}}
\newcommand{\paperref}[1]{\ensuremath{(#1)_{\text{\scriptsize\cite{Giesel:2017roz}}}}}
\newcommand{\gi}{^{\rm (gi)}}

\begin{document}
\title{{Dynamics of Dirac observables in canonical cosmological perturbation theory}}

\author{Kristina Giesel$^{1}$}
\thanks{kristina.giesel@gravity.fau.de}
\author{Parampreet Singh$^2$}
\thanks{psingh@lsu.edu}
\author{David Winnekens$^1$}
\thanks{david.winnekens@fau.de}
\affiliation{$^1$ Institute for Quantum Gravity, FAU Erlangen -- N\"urnberg,\\
Staudtstr. 7, 91052 Erlangen, Germany}
\affiliation{$^2$ Department of Physics and Astronomy,\\
Louisiana State University, Baton Rouge, LA 70803, U.S.A.}
\begin{abstract}
The relational formalism based on geometrical clocks and Dirac observables in linearized canonical cosmological perturbation theory is used to introduce an efficient method to find evolution equations for gauge invariant variables. Our method generalizes an existing technique by Pons, Salisbury and Sundermeyer \cite{Pons3, Pons4} to relate the evolution of gauge invariant observables with the one of gauge variant quantities, and is applied as a demonstration for the longitudinal and spatially flat gauges. Gauge invariant evolution equations for the Bardeen potential and the Mukhanov-Sasaki variable are derived in the extended ADM phase space. Our method establishes a full agreement at the dynamical level between the canonical and conventional cosmological perturbation theory at the linear order using Dirac observables. 
\end{abstract}

\maketitle

\section{Introduction}

Cosmological perturbation theory provides one of the most important avenues to test viability of cosmological models in general relativity and modified theories of gravity in the very early universe. While the standard approach based on perturbing the Einstein field equations over a background to the linear order \cite{bardeen,kodama-sasaki,Mukhanov} and the covariant approach \cite{Ellis:1989jt} are extensively developed, comparatively only little work has been done in the canonical framework. Partly, the reason of this gap can be attributed to the necessary 3+1 split in the canonical picture and along with this the non-trivial relation of gauge transformations in the Lagrangian and canonical framework to show the equivalence with the conventional approaches to perturbation theory. However, a Hamiltonian based cosmological perturbation theory becomes essential if one wishes to reliably understand effects of canonical quantization such as the way quantum gravitational modifications leave their signatures in cosmic microwave background like, for instance, addressed in the context of quantum cosmological models.
In addition, a pertinent question is whether cosmological perturbation theory can be formulated in canonical approach at an equivalent level to the conventional approaches.

While Langlois' seminal paper  obtained a phase space formulation of the Mukhanov-Sasaki variable \cite{Langlois} and its dynamics\footnote{A path integral formulation of cosmological perturbation theory in reduced phase space was presented in \cite{Anderegg:1994xq}.}, and has been a guiding light for investigations aiming to link canonical approaches with astronomical observations, it is important to note that being based on the ADM phase space with lapse and shift treated as Lagrange multipliers\footnote{This phase space is often referred to as the reduced ADM phase space.}, it leaves many fundamental questions open. 
An example of such a question is the way different gauges are treated, and if they affect any physical implications, particularly in modified theories of gravity and quantum gravity. For the latter, answering such questions becomes more complex since one has to simultaneously address the problem of time for example by invoking relational formalism. Note that an analysis based on reduced ADM phase space does not allow a full understanding of gauges such as the longitudinal, synchronous and comoving gauges involving lapse and/or shift perturbations in the context of the relational formalism, where gauge invariant quantities are constructed via an observable map on phase space that needs to be applied also to the lapse and shift degrees of freedom for these gauges. The extended phase space allows to implement diffeomorphisms that act on all gravitational degrees of freedom and not only the spatial ones as in the reduced ADM phase space. Working with an extended phase space makes it easier to bridge contact to the Lagrangian framework widely used in cosmological perturbation theory. In this sense, there is a gauge restriction in the reduced phase space approach because even though the Mukhanov-Sasaki variable is gauge invariant, it has a natural interpretation only in the spatially flat and uniform field gauges. Due to these kind of limitations of the reduced canonical phase space framework, canonical cosmological perturbation theory was never brought to the same stage as the standard approach.  The role of gauge choices, which in particular can play an important role in quantum theory, associated gauge invariant variables and their evolution equations in the canonical theory thus necessitate going beyond the reduced ADM phase space setting. 

To fill these gaps between the standard and canonical approaches to cosmological perturbation theory and to lay a platform to extract reliable predictions using the canonical approach, an extended ADM phase space formulation where lapse and shift are treated as dynamical phase space variables was introduced in \cite{Giesel:2017roz,Giesel:2018opa}. These works provided a first  formulation of canonical cosmological perturbation theory in extended phase space, not only opening  a window to establish the equivalence with the Lagrangian based perturbation theory but laying down a platform for further investigations such as higher order perturbation theory in the canonical setting. Based on the relational formalism, the essential idea involves choosing an appropriate set of reference fields, called clocks, which are compatible with the  Hamiltonian and spatial diffeomorphism constraints in general relativity. Using the observable map, gauge invariant extensions (so-called Dirac observables) of the metric and matter perturbations as well as their conjugate momenta can be obtained. These Dirac observables that are tied to the choice of clocks, which again carries over to a choice of gauge fixing conditions, naturally select the gauge invariant variables for the respective gauges, such as the Bardeen potentials for the longitudinal gauge and the Mukhanov-Sasaki variable for the spatially flat gauge, without re-coursing to look for appropriate combinations as in the conventional approaches. The method is straightforward and results in novel insights such as on the deeper relationship between the choice of clocks, gauge fixing conditions, Dirac observables and natural gauge invariant quantities in the cosmological perturbation theory. Note that if one does not wish to work in the context of the relational formalism and considers perturbations in the reduced ADM phase space along with perturbations of the Lagrange multipliers of lapse and shift, then at the level of the reduced phase space one can obtain the phase space dependent expressions for lapse and shift also using the stability condition for the gauge fixing condition (see for eg. \cite{Malkiewicz:2018ohk}). In the approach used in our work, this process is automatically and consistently built into the formalism. In our approach the observable map is needed for {\it{all}} degrees of freedom using the relational formalism, which necessitates  the usage of the extended phase space irrespective of the choice of gauge condition. 

The goal of this work is to carry forward our results in \cite{Giesel:2017roz,Giesel:2018opa} and find the corresponding evolution equations for the gauge invariant variables in canonical cosmological perturbation theory. This is achieved for the longitudinal and the spatially flat gauges in the extended phase space formulation. In our approach, we obtain the evolution equations by generalizing a lemma by Pons, Salisbury and Sundermeyer relating the evolution of gauge invariant observables and gauge variant quantities obtained earlier in \cite{Pons3,Pons4}. This generalization was necessary in order to incorporate the class of gauge fixing conditions relevant in cosmological perturbation theory.  This provides a new method to efficiently derive evolution equations for different gauge invariant variables in canonical cosmological perturbation theory.

Let us briefly recall the general setting of the clocks and relational observables approach and related work. In the application of the relational formalism \cite{RovelliPartial,RovelliObservable,Dittrich,Dittrich2} to the framework of cosmological perturbation theory, one approach is to start with full general relativity, choose appropriate clocks and construct manifestly gauge invariant observables. These are quantities that are invariant under gauge transformations up to any order. The associated gauge invariant evolution equations can then be derived and one obtains a gauge invariant version of the Einstein's equations. Given these evolution equations, one can compute their linearization and treat this as a possible setup for linearized cosmological perturbation theory. This has for instance been done in \cite{Giesel:2007wi, Giesel:2007wk} for dust matter clocks in the context of general relativity (see also \cite{Han:2015jsa} for an application to scalar-tensor theories and \cite{Giesel:2009jp} for LTB spacetimes). In all the cases, one first constructs non-linear gauge invariant quantities and afterwards considers perturbation theory. This is in contrast to an alternate route where one first considers perturbations of Einstein's equations up to a certain order and afterwards constructs quantities that are invariant under gauge transformations up to the order that one has chosen for the perturbation theory. 

One of the advantages of the first method is that once non-linear clocks are chosen, the construction of gauge invariant quantities of higher than linear order can be straightforwardly obtained, whereas in the second method in principle this has to be discussed order by order anew.  In the existing literature, the latter method was analyzed only in the context of matter reference fields by introducing additional dust or scalar field degrees of freedom to general relativity. An attempt to derive the gauge invariant sector with geometrical clocks, which are chosen among the gravitational degrees of freedom  in the reduced phase space where lapse and shift are treated as Lagrange multipliers, was performed for Ashtekar-Barbero variables \cite{Dittrich-Tambornino2}, albeit only for the longitudinal gauge. In a more recent work presented in our companion papers \cite{Giesel:2017roz, Giesel:2018opa}, several gauges relevant for cosmological perturbation theory have been considered in the extended ADM phase space. There, the authors also follow the second approach and show that the construction of the usual gauge invariant quantities such as the Bardeen potentials and the Mukhanov-Sasaki variable can be embedded into  language of the relational framework using the extended ADM phase space formulation that includes lapse and shift  as dynamical variables. However, non-linear geometrical clocks in this setting have not been considered in the literature so far, presumably because the resulting algebra of the observables is in general more complicated than the associated one for the gauge variant quantities. As a consequence, also their resulting quantization complicates a relevant aspect in a reduced phase space formulation of quantum gravity, which is for instance discussed in \cite{Giesel:2007wn,Domagala:2010bm}. The work in \cite{Giesel:2017roz,Giesel:2018opa} can thus be viewed  as a first step towards the usage of non-linear geometrical clocks because the results therein can provide hints on the suitable choice of non-linear geometrical clocks. 

In the previous works \cite{Giesel:2017roz,Giesel:2018opa}, we covered the first two main steps -- identifying appropriate clocks, and using the observable map to find the corresponding gauge invariant quantities for 5 common gauges in cosmological perturbation theory. However, the associated evolution equations were not analyzed in detail. As noted above, in this work, we fill this gap with the standard perturbation theory for the longitudinal and spatially flat gauge and show that the relational framework provides a convenient technique to compute these equations in certain aspects more efficiently than in the usual Lagrangian approach. Furthermore, this will also provide an additional verification that even at the dynamical level, the results derived in linearized cosmological perturbation theory in the Lagrangian framework can fully be rediscovered in a phase space formulation in terms of Dirac observables using the relational framework. For showing this, we have to generalize earlier results in \cite{Pons3,Pons4,Giesel:2007wi,Giesel:2007wk} because the gauge fixing conditions that are needed  to be used in \cite{Giesel:2017roz,Giesel:2018opa} to construct quantities like the Mukhanov-Sasaki variable and the Bardeen potentials are in some aspects more general than the ones considered in \cite{Pons3,Pons4,Giesel:2007wi,Giesel:2007wk}. In addition, we have to carefully analyze whether the properties of the evolution of the observables holds also in the context of less specialized gauge fixing conditions. 

The plan of our manuscript is as follows. Sec. \ref{Sec:EvolEqns} consists of two parts, where in the first part we summarize the 
general setup on the choice of clocks and construction of relational observables 
 in the setting of linear cosmological perturbation theory. We follow the conventions in \cite{Giesel:2017roz,Giesel:2018opa}, which we refer to the reader for further details.  Section \ref{Sec:GenEvolEq} provides the theoretical background and proofs to formulate the evolution equations of the observables such as the Bardeen potential and the Mukhanov-Sasaki variable completely at the gauge invariant level and in the context of the relation formalism. In this part, we generalize results of \cite{Pons4,Giesel:2007wi} that relate time evolution of gauge invariant observables with gauge variant quantities to accommodate cosmological gauge fixing conditions considered in this work. The new result of section \ref{Sec:GenEvolEq} is a generalization of a lemma that allows us to find in a straightforward way evolution equations for the Bardeen potential and the Mukhanov-Sasaki variable. Readers who are mainly 
interested in applications of this lemma to cosmological perturbation theory can directly skip to section \ref{Sec:Applic}.  In this section, this lemma is applied to derive Hamilton's equations for gauge invariant observables in longitudinal and spatially flat gauges, which immediately yield the evolution equations for both the Bardeen potential and the Mukhanov-Sasaki variable.  Finally, in  section \ref{Sec:SummConcl}, we summarize, conclude and discuss future applications of the work presented in this article. Various longer proofs and computations are included in appendices A--D. In particular, in appendix D we discuss the way how the weakly commuting property of Dirac observables with all constraints in the full non-linear theory holds also in the perturbation theory. 

\section{Evolution equations for gauge invariant variables}
\label{Sec:EvolEqns}
In this section, we will start with a very brief summary of the results in \cite{Giesel:2017roz,Giesel:2018opa} on clocks and Dirac observables, which will be used as an input for the analysis performed in this article. In the second part of this section, we present a general discussion on the way evolution equations for the relevant observables can be formulated at the gauge invariant level. In this part, we present a lemma generalizing results of Pons, Salisbury and Sundermeyer \cite{Pons3,Pons4} relating the dynamics of gauge invariant observables with the one of gauge variant quantities. 

\subsection{Summary of the choice of clocks and construction of observables in the relational framework}
We work in the extended ADM phase space of general relativity minimally coupled to a scalar field. Therefore, our gauge variant degrees of freedom are given by
\begin{equation}
(q_{ab},P^{ab}),\quad (N,\Pi),\quad (N^a,\Pi_a),\quad (\varphi,\pi_\varphi),	
\end{equation}
where $q_{ab}$ denotes the ADM metric, $N$ the lapse function, $N^a$ the shift vector and $\varphi$ the scalar field, and we have introduced their corresponding conjugate momenta. In the extended ADM phase space, we have 8 constraints of which four are primary -- $\Pi$ and $\Pi^a$ -- and four are secondary constraints -- the Hamiltonian $C^{\rm tot}$ and the spatial diffeomorphism constraint $C^{\rm tot}_a$:
\begin{equation}
\Pi\approx 0,\quad \Pi^a\approx 0,\quad
C^{\rm tot}=C^{\rm geo}+C^\varphi\approx 0,\quad
C_a^{\rm tot}=C_a^{\rm geo}+C_a^\varphi\approx 0	.
\end{equation}
Here, we denoted the gravitational and scalar field contributions to the Hamiltonian constraint by $C^{\rm geo}$ and $C^\varphi$ respectively, while $C^{\rm tot}_a$ labels the total constraint. The canonical Hamiltonian generating the equations of motions for the gauge variant quantities has the following  form:
\begin{equation}
\label{eq:CanHam}
H_{\rm can}=\int \d^3x\left(NC^{\rm tot}+N^aC_a^{\rm tot}+\lambda\Pi+\lambda_a\Pi^a\right)
=\int \d^3x\left(N^\mu C^{\rm tot}_\mu+\lambda^\mu\Pi_\mu\right)	~.
\end{equation}
In the Hamiltonian framework, Einstein's field equations are encoded in the 8 constraints as well as the first order Hamiltonian equations for the elementary phase space variables, that is,
\begin{equation}
\dot{q}_{ab}=\{q_{ab},H_{\rm can}\},\quad
\dot{P}^{ab}=\{P^{ab},H_{\rm can}\},	
\end{equation}
and similarly for the remaining gauge variant degrees of freedom.

Next, we consider linear perturbations around a spatially flat FLRW background. In this case, the spacetime metric is given by
\begin{equation}
    \d s^2 = - \bar N^2(t) \d t^2 + \bar q_{ab} \d x^a \d x^b = - \bar N^2(t) \d t^2 + A(t) \delta_{ab} \d x^a \d x^b, ~~~ a,b=1,2,3. 
\end{equation}
Here,  $A(t) \coloneqq a^2(t)$, where $a(t)$ is the scale factor of the universe. For this spacetime metric, 
the elementary variables in the extended phase space take the form
\begin{align}
\label{eq:FLRWPert}
q_{ab}(\vec{x},t) &= A(t)\delta_{ab} + \delta q_{ab}(\vec{x},t), & P^{ab}(\vec{x},t) &= \tilde{P}(t)\delta^{ab} + \delta P^{ab}(\vec{x},t), \nonumber \\
N(\vec{x},t) &= \bar{N}(t) + \delta N(\vec{x},t), &  \Pi(\vec{x},t) &= \delta \Pi(\vec{x},t), \nonumber \\
N^a(\vec{x},t) &=  \delta N^a(\vec{x},t), & \Pi_a(\vec{x},t) &= \delta \Pi_a(\vec{x},t),\nonumber\\
\varphi(\vec{x},t) &= \bar{\varphi}(t) + \delta \varphi(\vec{x},t), &
\pi_\varphi(\vec{x},t) &= \bar{\pi}_\varphi(t)+\delta \pi_\varphi(\vec{x},t),
\end{align}
where we denote background quantities with a bar and linear perturbations with a $\delta$. Here, $P_A=3\tilde P$ captures the momentum conjugate to $A$, where $\tilde{P}$ is  given by  
\begin{equation}
\tilde P = -\frac{\dot A}{\Nb\sqrt{A}} ~.
\end{equation}
Similarly, we can rewrite the secondary constraints in terms of their perturbations as
\begin{equation*}
C^{\rm tot}(\vec{x},t)=\delta C^{\rm geo}(\vec{x},t)+\delta C^{\varphi}(\vec{x},t),\quad
C_a^{\rm tot}(\vec{x},t)=\delta C_a^{\rm geo}(\vec{x},t)+\delta C_a^{\varphi}(\vec{x},t),\quad
\end{equation*}
where we took into account that the background constraints are assumed to vanish because the background is a solution of Einstein's field equations. Since scalar, vector and tensor degrees of freedom decouple at the linear order, we can express the perturbations in the lapse, shift and spatial metric using  scalar-vector-tensor decomposition:
\begin{align}
\delta N &= \bar{N}\phi,\\
\delta N^a &= B^{,a} + S^a ~~~~~ \mathrm{and}\nonumber \\
\delta q_{ab} &= 2A\left( \psi \delta_{ab} + E_{,<ab>} + F_{(a,b)} + \tfrac{1}{2}h^{TT}_{ab} \right) .
\end{align}
Likewise, we can perform a similar scalar-vector-tensor decomposition for the conjugate momenta to obtain
\begin{eqnarray}
\delta P^{ab}& =& 2\tilde{P}\left( p_\psi\delta^{ab} + p_E^{,<ab>} + p_F^{(a,b)} + \tfrac{1}{2}p_{h^{TT}}^{ab} \right), \\
p_\phi &=&\frac{1}{\overline{N}}\delta\Pi, \quad p_B =\Delta^{-1}\partial_a\delta \Pi^a\eqqcolon\delta\hat{\Pi} \ {\rm and}\nonumber\\
 p_{S^a}&=&\delta\Pi^a-\partial^a(\Delta^{-1}\partial_b\delta \Pi^b)\eqqcolon\delta\Pi^a_\perp ,
\end{eqnarray}
where $\Delta^{-1}$ denotes the Green's function of the Laplacian with respect to the metric $\delta_{ab}$ and, as before, $\delta\Pi$ and $\delta\vec{\Pi}$ denote the conjugate momenta of the perturbed lapse function $\delta N$ and the perturbed shift vector $\delta\vec{N}$ respectively.

The first step in our strategy consists of choosing an appropriate set of clocks consistent with the constraints. In the context of cosmological perturbation theory, we want to construct quantities that are invariant under linearized diffeomorphisms.  Hence, we want to construct specific combinations of the elementary variables on the linearized phase space that are gauge invariant up to terms that are of second or higher order. Since the scalar, vector and tensor sector decouple, we can do this separately for each sector. Further, we can neglect the tensor sector because their variables are already gauge invariant.  The way non-linear diffeomorphisms can be implemented on the extended ADM phase space was introduced in \cite{Pons1,Pons2,Pons3,Pons4}. An application of their formalism to linearized cosmological perturbation theory was studied in \cite{Giesel:2017roz,Giesel:2018opa}. It turns out that implementing our first step amounts to  the choice of four clocks on the linearized phase space denoted by $\delta T^\mu$, with $\mu=0,1,2,3$, where $\delta T^0$ is the temporal clock associated with temporal diffeomorphism and $\delta T^a$ correspond to the three clocks for the spatially diffeomorphism constraints. In the relational formalism, choosing $\delta T^\mu$ can be understood as a choice of physical temporal and spatial coordinates associated with the reference fields. In order to apply them in the scalar and vector sector, we use their corresponding projections (see \cite{Giesel:2018opa} for further details). As analyzed in detail in \cite{Giesel:2018opa}, there exists a natural choice for the set of clocks for each common gauge. Therefore, these sets of clocks that we choose are {\it{different}} for the longitudinal and spatially flat gauges. 

Given these sets of clocks, we can perform the second step of our strategy to obtain relevant observables. This is achieved by applying the observable map on the linearized phase space. The most general formula for such observables is given by
\begin{align}
\label{eq:pertobs2}
\delta \mathcal{O}_{f,T}[\tau] &= \delta f + \int\mathrm{d}^3y \left[ \delta \dot{G}^\mu(y)\overline{\{ f,\tilde{\Pi}_\mu(y) \}} + \delta T^\mu(y)\overline{\{ f,\tilde{\tilde{C}}_\mu(y) \}} \right] \nonumber \\
~ &\approx \delta f + \int\mathrm{d}^3y \int\mathrm{d}^3z \bar{\mathcal{B}}^\nu_\mu(z,y) \left[ \delta \dot{G}^\mu(y)\overline{\{ f,\Pi_\nu(z) \}} - \delta G^\mu(y) \left( \overline{\{ f,C_\nu(z) \}} \right. \right. \nonumber \\
~ &\hphantom{\approx}\ + \left. \left. \int\mathrm{d}^3w \int\mathrm{d}^3v~ \bar{\mathcal{B}}^\rho_\sigma(w,v) \overline{\{ \dot{T}^\sigma(v),C_\nu(z) \}}~ \overline{\{ f, \Pi_\rho(w) \}} \right) \right].
\end{align}
Let us explain our notation in detail. The function $f$ is the quantity for which we want to construct gauge invariant extensions. $\delta G^\mu$ denotes the perturbation of the four gauge fixing conditions. Often, gauge fixing conditions of the form $G^\mu=\tau^\mu-T^\mu$ are considered, where $\tau^\mu$ is in general a function on spacetime. Hence, parametrized by $\tau^\mu$, we obtain a family of Dirac observables for each choice of $\tau^\mu$. As we will discuss below, we need to allow more general gauge fixing conditions for our work in the context of cosmological perturbation theory. The constraints $\Pi_\mu=(\Pi,\Pi_a)$ and $C_\mu=(C^{\rm tot},C_a^{\rm tot})$ encode the set of primary and secondary constraints. The tildes on top of constraints refer to certain linear combinations of them that define the same constraint surface. The quantity ${\cal B}^\mu_\nu$ denotes a matrix element of the inverse matrix of ${\cal A}^\mu_\nu\coloneqq \{T^\mu,C_\nu\}$. Finally, the background quantities are indicated with a bar on their top. Note that the existence of the inverse matrix is necessary, which is a condition the chosen clocks must satisfy. The time derivatives of the gauge fixing conditions and clocks that occur in the observable formula are due to the stability conditions of the gauge fixing conditions. A detailed derivation of this formula can be found in \cite{Giesel:2018opa}. Moreover, it is also analyzed there what kind of clocks and gauge fixing conditions respectively have to be chosen such that the observable map in \eqref{eq:pertobs2} yields automatically the common gauge invariant quantities used in cosmological perturbations theory. These are the Bardeen potentials for the longitudinal gauge and the Mukhanov-Sasaki variable for the spatially flat gauge. Here, we will just list the results of \cite{Giesel:2018opa} that can be found in table \ref{tab:clockchoices1}. In this table, for each projected elementary gauge variant quantity in the scalar and vector sector, the corresponding Dirac observables are listed for the longitudinal and spatially flat gauge, together with the choice of clocks relevant for the individual gauges. We take this as the starting point for this paper, where the evolution equations for these observables are derived and analyzed. 

\subsection{The evolution equations for the observables}
\label{Sec:GenEvolEq}
The evolution of the gauge variant quantities are generated by the canonical Hamiltonian in (\ref{eq:CanHam}). In this subsection, we want to carefully analyze how we can formulate the equations of motion at the level of the observables. 
For this purpose, we have to generalize results presented in \cite{Pons3,Pons4} (and also \cite{Giesel:2007wi}) to a class of gauge fixing conditions that are relevant in the context of cosmological perturbation theory. Our discussion will involve generic gauge fixing conditions, and, in the next section, we will apply our results to the longitudinal as well as the spatially flat gauge, for which the associated gauge fixing conditions fall in the class considered here. 

The work in \cite{Giesel:2007wi,Giesel:2007wk} considered coordinate gauge fixing conditions of the form
\begin{equation}
\label{eq:LinGF}
G^\mu=\tau^\mu-T^\mu \quad \quad \mathrm{with} \quad \quad \mu=0,1,2,3,	
\end{equation}
where $\tau^\mu$ is a generic spacetime function and the clocks $T^\mu$ are linear in one of the configuration variables. In particular, $T^\mu$ are chosen as the four dust scalar fields that are coupled to general relativity in addition to a minimally coupled scalar field. It should be noted that the gauge fixing conditions chosen in \cite{Pons4,Pons3} are a special case of the ones in (\ref{eq:LinGF}) because the former consider $\tau^\mu=x^\mu$ and  also four scalar fields as clocks, whose dynamics, however, is not specified. 

The class of gauge fixing conditions we want to analyze here is of the form
\begin{equation}
\label{eq:GenGF}
G^\mu=-T^\mu(q^A,p_A,x^\mu) \quad \quad \mathrm{with} \quad \quad \mu=0,1,2,3,	
\end{equation}
where we introduced a compact notation for the configuration variables $q^A=(q_{ab},\varphi)$ and their conjugate momenta 
$p_A=(P^{ab},\pi_\varphi)$. Here, for $T^\mu$, we assume a generic dependence on phase space variables that do not involve the lapse and shift degrees of freedom and only a possible dependence on the temporal and spatial coordinates\footnote{Note that this restriction does not exclude gauges where, in the Lagrangian framework, the corresponding lapse or shift degrees of freedom are gauge fixed such as, for instance, in the longitudinal gauge. As shown in \cite{Giesel:2018opa}, these gauges can be reproduced in this framework.}. For our applications in the next section, $T^\mu(q^A,p_A,x^\mu)$ will always be a linear function of the phase space variables, but not necessarily only of one variable. Further, its dependence is not restricted to configuration variables only. Moreover, it will depend explicitly only on the temporal coordinate. In the case of cosmological perturbation theory for the gauges considered here, the time dependence is due to involved background quantities associated with the background FLRW solution. 

In general relativity, we need to choose four independent gauge fixing conditions, whose stability requirement will induce specific relations among lapse, shift and the remaining variables on phase space. In the reduced ADM phase space, the Lagrange multipliers for lapse and shift will be fixed by the stability of the gauge fixing conditions and thus become phase space dependent. On the other hand, in  case of the extended ADM phase space, lapse and shift are among the phase space variables -- and the stability of the gauge fixing conditions induces relations among the phase space variables. A crucial observation is that at least one of the gauge fixing conditions needs to be explicitly time dependent as, otherwise, lapse and shift are fixed to be trivial function on phase space. This can be easily seen from the stability of the gauge fixing condition. We have
\begin{eqnarray}
\frac{\d G^\mu}{\d t}(x)&=&\{G^\mu,H_{\rm can}\}+\frac{\partial G^\mu}{\partial t}(x)\nonumber \\
&=&
\frac{\partial G^\mu}{\partial t}(x)+\int\limits \d^3y N^\nu(y)\{G^\mu(x),C_\nu(y)\}\nonumber \\
&\approx&\frac{\partial G^\mu}{\partial t}(x)-\int\limits \d^3y N^\nu(y){\cal A}^\mu_\nu(x,y)\nonumber\\
&\ \ \stackrel{!}{=}0 ,
\end{eqnarray}
where we used in the second line that $G^\mu$ does not depend on lapse and shift degrees of freedom and in the third line that $\{G^\mu(x),C_\nu(y)\}\approx -{\cal A}^\mu_\nu(x,y)$, where the weak equality here is with respect to the secondary constraints. From the stability requirement, we immediately obtain
\begin{equation}
\label{eq:GmuStab}
N^\mu(x)\approx \int\limits \d^3y \frac{\partial G^\nu}{\partial t}(y){\cal B}^\mu_\nu(x,y) \quad {\rm for}\quad \mu=0,\ldots,3 .	
\end{equation}
Note that all $N^\mu$ vanish if we choose all $G^\mu$ with no explicit time dependence. This is ensured in  the work in \cite{Pons4,Giesel:2007wi,Giesel:2007wk}  because at least $\tau^0$ is assumed to depend on time. In our case, the explicit time dependence is encoded in the function $T^\mu(t)$, which is always chosen to be a non-trivial function at least for one of the $\mu$'s. 

Now, if we use gauge fixing conditions of the form in (\ref{eq:LinGF}), as has been done in \cite{Pons3,Pons4,Giesel:2007wi,Giesel:2007wk}, we can interpret the dynamics of the observables as an evolution with respect to the parameter $\tau^0$, which is the value the chosen clock $T^0$ takes.  However, using the gauge fixing conditions in (\ref{eq:GenGF}), this interpretation seems to be lost. This is because in order to obtain  common gauge invariant variables such as the Bardeen potential as well as the Mukhanov-Sasaki variable, specific components of the perturbed metric degrees of freedom are set to zero, and they are exactly those from which we build our clocks on the phase space. As we will show below, due to the dependence of at least one of the functions $T^\mu$ on the temporal coordinate, it is nevertheless possible to define a sensible notion of evolution for the observables. We also discuss the way these observables can be interpreted. We will analyze this at the full non-linear level of the observables and later simply specialize this to the linear order relevant for linearized cosmological perturbation theory. The way the condition for Dirac observables as well as the formulation of their dynamics can be carried over to perturbation theory has been in detail discussed in appendix \ref{sec:DiracObsPert} building on former work in \cite{Giesel:2007wi,Giesel:2007wk}, where the notion of constants of motion in the context of perturbation theory has been analyzed. At the non-linear level, the observable of a given function $f$ on the extended phase space has the following form:
\begin{equation}
\label{eq:obsGR}
\mathcal{O}_{f,T}=f+ \sum\limits_{n=1}^{\infty} \frac{1}{n!} \int\mathrm{d}^3y_1...\int\mathrm{d}^3y_n~ \mathfrak{G}^{I_1}(y_1)...\mathfrak{G}^{I_n}(y_n) \{...\{ f,\tilde{\mathfrak{C}}_{I_1}(y_1) \},...\tilde{\mathfrak{C}}_{I_n}(y_n) \} .
\end{equation}
Here, we introduced the notation $\mathfrak{G}^I \coloneqq (G^\mu,G^{(2)\mu})$ and $\tilde{\mathfrak{C}}_I(x) \coloneqq (\tilde{\tilde{C}}_\mu,\tilde{\Pi}_\mu)$, where $G^{(2)\mu}$ denote the conditions obtained from the stability requirement of the gauge fixing conditions $G^\mu$, while $\tilde{\tilde{C}}_\mu$ and $\tilde{\Pi}_\mu$ define specific abelianized versions of the secondary and primary constraints respectively that define the same constraint hypersurface as the original ones. To obtain them, we need to consider the Poisson brackets among the individual constraints:
\begin{equation}
\label{eq:AMatrix}
\mathfrak{A}^I_J \coloneqq -\{ \mathfrak{G}^I,\mathfrak{C}_J \} = - \left[ \begin{matrix}
\mathcal{A}^\mu_\nu & 0 \\
\{ \dot{T}^\mu,C_\nu \} & \mathcal{A}^\mu_\nu
\end{matrix} \right].
\end{equation}
Note that we have used the identity $\dot{T}^\mu=\partial_tT^\mu + \int\mathrm{d}^3x~ \mathcal{A}^\mu_\nu(\cdot,x) N^\nu(x)$ for the last entry of the above matrix, and we further used the already introduced definition  $\mathcal{A}^\mu_\nu \coloneqq \{ T^\mu,C_\nu \}$. Let us denote the inverse of 
$\mathfrak{A}$ by $\mathfrak{B}$. As shown in \cite{Pons3}, $\mathfrak{B}$ can be easily computed and has the following form:
\begin{equation}
\label{eq:InverseAMatrix}
\mathfrak{B}^I_J = (\mathfrak{A}^{-1})^I_J = \left[ \begin{matrix}
\mathcal{B}^\mu_\nu & 0 \\
S^\mu_\nu & \mathcal{B}^\mu_\nu
\end{matrix} \right],
\end{equation}
with
\begin{equation}
S^\mu_\nu(x,y) = -\int\mathrm{d}^3z\int\mathrm{d}^3v~ \mathcal{B}^\mu_\rho(x,z)\mathcal{B}^\sigma_\nu(v,y)\{ \dot{T}^\rho(z),C_\sigma(v) \}.
\end{equation}
The abelianized constraints that enter the observable map can be constructed by using $\mathfrak{B}$, yielding
\begin{equation}
\tilde{\mathfrak{C}}_I(x) = \int\mathrm{d}^3y~\mathfrak{B}^J_I(y,x)\mathfrak{C}_J(y) ~.
\end{equation}
We define the equivalent abelian set of constraints by $\tilde{\mathfrak{C}}_I(x) \eqqcolon (\tilde{\tilde{C}}_\mu,\tilde{\Pi}_\mu)$. Using $\mathfrak{B}^I_J$, their form in terms of the original constraints $C_\mu$, $\Pi_\mu$ is explicitly given by
\begin{align}
\label{eq:RelationConst}
\tilde{\Pi}_\mu(x) &= \int\mathrm{d}^3y~\mathcal{B}^\nu_\mu(y,x)\Pi_\nu(y) ,  \nonumber \\
\tilde{\tilde{C}}_\mu(x) &= \int\mathrm{d}^3y~\mathcal{B}^\nu_\mu(y,x) \left[ C_\nu(y) - \int\mathrm{d}^3z\int\mathrm{d}^3v~ \mathcal{B}^\sigma_\rho(v,z)\{ \dot{T}^\rho(z),C_\nu(y) \}\Pi_\sigma(v) \right].
\end{align}
More details and a derivation of this formal power series for the observables can be found in \cite{Giesel:2017roz}.
Due to the gauge-fixing conditions and their temporal derivatives involved in $\mathfrak{G}^I$, the observable $\mathcal{O}_{f,T}$ depends explicitly on time and we can compute its total time derivative:
\begin{eqnarray}
\frac{\d \mathcal{O}_{f,T}}{\d t}
&=&
\frac{\partial \mathcal{O}_{f,T}}{\partial t}
+\{\mathcal{O}_{f,T},H_{\rm can}\}
\approx \frac{\partial \mathcal{O}_{f,T}}{\partial t},
\end{eqnarray}
where we used in the last step that, in the context of general relativity, all observables commute with $H_{\rm can}$ by construction. In appendix \ref{sec:DiracObsPert} we derive the corresponding equation in the context of perturbation theory, where a similar result holds. The equation above also motivates why we are choosing gauge fixing conditions that depend explicitly on time because the way the observables are constructed their explicit time dependence enters via the gauge fixing conditions.  In this work in particular, we want to prove the following lemma:
\begin{lemma}
\label{MainLemma}
For a gauge fixing condition of the form in (\ref{eq:GenGF}) that is allowed to depend on all phase variables except lapse function and shift vector  degrees of freedom, and that can have an explicit dependence on coordinates, and for which the associated matrix ${\cal A}^\mu_\nu\coloneqq \{T^\mu,C_\nu\}$ has the property that  $\frac{\partial {\cal A}^\mu_\nu}{\partial t}\approx 0$, the following result holds:
\begin{equation}
\frac{\d \mathcal{O}_{f,T}}{\d t}\approx \mathcal{O}_{\{f,H_{\rm can}\},T}.	
\end{equation}	
\end{lemma}
A similar property was proven in \cite{Pons4}, however, only  for a class of gauge fixing conditions $G^\mu=x^\mu-T^\mu$ that cannot be applied to the case of cosmological perturbation theory. As we will show below, the proof in \cite{Pons4} can be generalized but involves additional non-trivial steps which will be fulfilled in our analysis. Furthermore, this result allows to derive the evolution equations for the observables without explicitly knowing the physical Hamiltonian. This will be of advantage in the context of cosmological perturbation theory with geometrical clocks, as discussed in this work, where the derivation of the physical Hamiltonian is less straightforward than in models with matter clocks, as for instance in \cite{Giesel:2007wi,Giesel:2007wk,Giesel:2012rb,Domagala:2010bm,Giesel:2009jp,Han:2015jsa}.

 Let us briefly comment on our assumptions used in the lemma. First, we require the gauge fixing condition to be independent of lapse and shift degrees of freedom, as otherwise the zero submatrix in ${\mathfrak A}^J_K$ in (\ref{eq:AMatrix}) is non-vanishing and thus the entire observable map needs to be generalized. The second assumption, namely that $\frac{\partial {\cal A}^\mu_\nu}{\partial t}\approx 0$, is necessary in order to ensure that all additional terms that occur in the temporal derivative of the observable weakly vanish. Note that gauge fixing conditions of the form $G^\mu=x^\mu-T^\mu$ and $G^\mu=\tau^\mu-T^\mu$, where $T^\mu$ is one of the configuration variables, as used for instance in \cite{Pons3,Giesel:2007wi}, obviously satisfy $\partial_t {\cal A}^\mu_\nu=0$ and thus can be seen as a special subclass of the gauge fixing conditions considered here. Furthermore, it is also necessary to still be able to replace the gauge generator ${\mathfrak C}_I$ by the weakly equivalent one, shown in (\ref{eq:CundCPrime}) in the appendix where this aspect is also discussed in more detail. As we will show, from the assumption $\frac{\partial {\cal A}^\mu_\nu}{\partial t}\approx 0$, it automatically follows that $\frac{\partial {\mathfrak A}^I_J}{\partial t}\approx 0$. Thus, our  assumption is sufficiently strong for our purpose. This is presented in detail in the appendix (see the discussion in the paragraph below  (\ref{eq:PartialPB})).

We will prove lemma \ref{MainLemma} by means of several intermediate results. 
As a first step, we want to have an explicit form for the time derivative of the observables $\mathcal{O}_{f,T}$. Using the expression of the observable in (\ref{eq:obsGR}), we get for its time derivative
\begin{align}
\label{eq:dOdt}
\frac{\d \mathcal{O}_{f,T}}{\d t}
&\approx
\sum\limits_{n=1}^{\infty} \frac{1}{(n-1)!} \int\mathrm{d}^3y_1...\int\mathrm{d}^3y_n \nonumber\\
& \qquad\! \mathfrak{G}^{I_1}(y_1)...\mathfrak{G}^{I_{n-1}}(y_{n-1})\frac{\partial\mathfrak{G}^{I_n}}{\partial t}(y_n)  \{...\{ f,\tilde{\mathfrak{C}}_{I_1}(y_1) \},...\tilde{\mathfrak{C}}_{I_n}(y_n) \}\nonumber\\
&=
\sum\limits_{n=0}^{\infty} \frac{1}{n!} \int\mathrm{d}^3y_1...\int\mathrm{d}^3y_n\int\mathrm{d}^3y_{n+1}~
 \nonumber\\
&\qquad\! \mathfrak{G}^{I_1}(y_1)...\mathfrak{G}^{I_{n}}(y_{n})\frac{\partial\mathfrak{G}^{I_{n+1}}}{\partial t}(y_{n+1})  \{...\{ f,\tilde{\mathfrak{C}}_{I_1}(y_1) \},...\tilde{\mathfrak{C}}_{I_n}(y_n) \},\tilde{\mathfrak{C}}_{I_{n+1}}(y_{n+1})\} \nonumber\\
&\approx
\sum\limits_{n=0}^{\infty} \frac{1}{n!} \int\mathrm{d}^3y_1...\int\mathrm{d}^3y_n\int\mathrm{d}^3y_{n+1}~
 \nonumber\\
& \qquad\! \mathfrak{G}^{I_1}(y_1)...\mathfrak{G}^{I_{n}}(y_{n})\frac{\partial\mathfrak{G}^{I_{n+1}}}{\partial t}(y_{n+1})  \{...\{ f,\tilde{\mathfrak{C}}_{I_{n+1}}(y_{n+1}) \},\tilde{\mathfrak{C}}_{I_1}(y_1)\},...\tilde{\mathfrak{C}}_{I_n}(y_n) \} ,
\end{align}
where we used in the last line that the weakly abelianized constraints weakly commute, and, hence, we can reshuffle the order of their associated Hamiltonian vector fields. Furthermore, for the first weak equivalence in the first line, we took into account the assumption that $\partial_t {\cal A}^\mu_\nu\approx 0$. As shown in the appendix, this carries over to $\partial_t {\mathfrak A}^\mu_\nu\approx 0$ and consequently $\partial_t{\mathfrak B}^\mu_\nu\approx 0$ so that all possible contributions that could in principle occur when the partial temporal derivative acts on the iterated Poisson bracket weakly vanish\footnote{Furthermore, we checked that we can replace the $\tilde{\mathfrak C}_I$ by a weakly equivalent expression that differs only by terms that are quadratic in the primary and secondary constraints $\tilde{\tilde{C}}_\mu, \tilde{\Pi}_\mu$ without changing the final result. A detailed discussion on the point can be found in section \ref{Sec:GenProof} in the appendix.}. 

Now, let us consider the partial time derivative of the gauge fixing conditions $\mathfrak{G}^{I}$ in more detail. We have
\begin{eqnarray}
\label{eq:partGI}
\frac{\partial\mathfrak{G}^{I}}{\partial t}
=\left(\frac{\partial G^\mu}{\partial t},\frac{\partial}{\partial t}\frac{\d G^\mu}{\d t}\right) ~.
\end{eqnarray}
The second entry, $\frac{\partial}{\partial t}\frac{\d G^\mu}{\d t}$, yields
\begin{eqnarray}
\frac{\partial}{\partial t}\frac{\d G^\mu}{\d t}(x)
& \approx &
-\{\frac{\d G^\mu}{\d t}(x),H_{\rm can}\}=\{\frac{\d T^\mu}{\d t}(x),H_{\rm can}\}\nonumber \\
&=&
\{\frac{\partial T^\mu}{\partial t}(x),H_{\rm can}\}
+\int \d^3y\{\{T^\mu(x),N^\rho(y)C_\rho(y)\},H_{\rm can}\}
\nonumber \\
&=&
\int \d^3y \{\frac{\partial T^\mu}{\partial t}(x),N^\rho(y)C_\rho(y)\}
+\int \d^3y\{T^\mu(x),C_\rho(y)\}\{N^\rho(y),H_{\rm can}\}
\nonumber\\
&&+\int \d^3yN^\rho(y)\{\{T^\mu(x),C_\rho(y)\},H_{\rm can}\}
\nonumber\\
&=&
\int \d^3y \{\frac{\partial T^\mu}{\partial t}(x),N^\rho(y)C_\rho(y)\}
+\int \d^3y \lambda^\rho(y){\cal A}^\mu_\rho(x,y)
\nonumber \\
&& +\int \d^3yN^\rho(y)\{\{T^\mu(x),C_\rho(y)\},H_{\rm can}\}.
\end{eqnarray}
Here, we used in the first line the stability of the gauge fixing condition, in the third line that the gauge fixing condition does not depend on lapse and shift degrees of freedom and in the last step the equations of motions for $N^\mu$, that is lapse and shift on the extended phase space. Now, for the last term in the last line, we use the Jacobi identity leading to
\begin{equation}
\{\{T^\mu(x),C_\rho(y)\},H_{\rm can}\}
\approx -\{\{H_{\rm can},T^\mu(x)\},	C_\rho(y)\}
=\{\{T^\mu(x),H_{\rm can}\},C_\rho(y)\} .
\end{equation}
We combine this term with the one that involves a temporal partial derivative of $T^\mu$ and get a total time derivative, which we denote by $\dot{T}^\mu$:
\begin{eqnarray}
\frac{\partial}{\partial t}\frac{\d G^\mu}{\d t}(x)
& \approx &
\int \d^3y \lambda^\rho(y){\cal A}^\mu_\rho(x,y)
+\int \d^3yN^\rho(y)\{\dot{T}^\mu(x),C_{\rho}(y)\}
\nonumber \\
& \approx &
\int \d^3y \lambda^\rho(y){\cal A}^\mu_\rho(x,y)
+\int \d^3y\int \d^3z \frac{\partial G^\nu}{\partial t}(z){\cal B}^\rho_\nu(y,z)\{\dot{T}^\mu(x),C_\rho(y)\},\nonumber\\
&&
\end{eqnarray}
where we used the stability of the gauge fixing condition, (\ref{eq:GmuStab}), in the last step. This equation can be easily solved for $\lambda^\mu$ yielding the following expression for the Lagrange multiplier:
\begin{align}
\label{eq:FixLambda}
\lambda^\mu(x) &= \int \d^3y\frac{\partial}{\partial t}\frac{\d G^\nu}{\d t}(y){\cal B}^\mu_\nu(x,y) \nonumber\\
& \quad -\int \d^3y\int \d^3u\int \d^3v\frac{\partial G^\nu}{\partial t}(v){\cal B}^\rho_\nu(u,v)
{\cal B}^\mu_\lambda(x,y)\{\dot{T}^\lambda(y),C_\rho(u)\}.						
\end{align}
Thus, with (\ref{eq:GmuStab}) and (\ref{eq:FixLambda}), we have determined $N^\mu$ as well as the Lagrange multipliers $\lambda^\mu$ in terms of the derivatives of the gauge-fixing conditions. These results will be reinserted into (\ref{eq:dOdt}), where the temporal partial derivatives of $\mathfrak{G}^{I}$ are involved. In order to prove lemma \ref{MainLemma}, we have to show that we can pull $\frac{\partial\mathfrak{G}^I}{\partial t}$ inside the $n$-fold iterated Poisson bracket. Of course, this yields several additional terms that in general will not vanish even weakly. However, for the class of gauge fixing conditions of the form (\ref{eq:GenGF}), we will show that all these extra terms vanish weakly. In the case of the gauge fixing conditions considered in \cite{Pons3,Pons4} and \cite{Giesel:2012rb,Giesel4} $\mathfrak{G}^{I}$ are phase space independent and hence lemma 1 is trivially satisfied. Note that in general this is no longer given for the generalized class of gauge fixing conditions considered in this work.

In the following, we will consider the specialized case where the function $f$ in $\mathcal{O}_{f,T}$ is assumed to be independent of the lapse and shift degrees of freedom. Afterwards, we will discuss the more general case in appendix \ref{Sec:GenProof}.

In the case where $f$ is a function on the reduced ADM-phase space and hence does not depend on lapse and shift or their conjugate momenta, the observable formula in $\eqref{eq:obsGR}$ simplifies to
\begin{equation}
\label{eq:obsSpec}
\mathcal{O}_{f,T}=f+ \sum\limits_{n=1}^{\infty} \frac{1}{n!} \int\mathrm{d}^3y_1...\int\mathrm{d}^3y_n~ G^{\mu_1}(y_1)...G^{\mu_n}(y_n) \{...\{ f,\tilde{C}_{\mu_1}(y_1) \},...\tilde{C}_{\mu_n}(y_n) \}. 
\end{equation}
In order to rewrite the iterated Poisson bracket in a more compact notation, we will introduce the Hamiltonian vector fields associated to $\tilde{C}_\mu$ denoted by $X_{\tilde{C}_\mu}\cdot f\coloneqq \{f,\tilde{C}_\mu\}$, with $\tilde{C}_\mu(x)=\int \d^3y {\cal B}^\nu_\mu(y,x)C_\nu(y)$. Then, we can rewrite the observable formula in (\ref{eq:obsSpec}) in the following way:
\begin{equation}
\label{eq:obsSpec2}
\mathcal{O}_{f,T}=f+\sum\limits_{n=1}^{\infty} \frac{1}{n!} \int\mathrm{d}^3y_1...\int\mathrm{d}^3y_n~ G^{\mu_1}(y_1)...G^{\mu_n}(y_n)X_{\tilde{C}_{\mu_n}(y_n)}\cdots X_{\tilde{C}_{\mu_1}(y_1)}\cdot f.
\end{equation}
Now, in the special case considered here and expressed in terms of the Hamiltonian vector fields, the time derivative of the observables reads
\begin{eqnarray}
\label{eq:TimeDerObsI}
\frac{\d \mathcal{O}_{f,T}}{\d t}&=&\sum\limits_{n=0}^{\infty} \frac{1}{n!} \int\mathrm{d}^3y_1...\int\mathrm{d}^3y_{n+1}~ G^{\mu_1}(y_1)...G^{\mu_n}(y_n)
\frac{\partial G^{\mu_{n+1}}}{\partial t}(y_{n+1})\nonumber \\
&& X_{\tilde{C}_{\mu_n}(y_n)}\cdots X_{\tilde{C}_{\mu_1}(y_1)}X_{\tilde{C}_{\mu_{n+1}}(y_{n+1})}\cdot f.
\end{eqnarray}
In order to be able to pull $\frac{\partial G^{\mu_{n+1}}}{\partial t}(y_{n+1})$ inside the iterated Poisson bracket as needed for proving lemma \ref{MainLemma}, we need one more lemma discussed below.\\
\begin{lemma}
\label{Lemma2}
For a gauge fixing condition of the form in (\ref{eq:GenGF}) that is allowed to depend on all phase variables except lapse function and shift vector degrees of freedom, and that can have an explicit dependence on coordinates, and for which the associated matrix ${\cal A}^\mu_\nu(x,y)\coloneqq -\{G^\mu(x),C_\nu(y)\}$ has the property that  $\frac{\partial {\cal A}^\mu_\nu}{\partial t}\approx 0$, we have for all $n\in\mathbf{N}$:
\begin{align}
\frac{\partial G^{\mu_{n+1}}}{\partial t}(y_{n+1})
 X_{\tilde{C}_{\mu_n}(y_n)}\cdots X_{\tilde{C}_{\mu_1}(y_1)}X_{\tilde{C}_{\mu_{n+1}}(y_{n+1})}\cdot f 
 & \nonumber \\
  \approx X_{\tilde{C}_{\mu_n}(y_n)}\cdots X_{\tilde{C}_{\mu_1}(y_1)}X_{\frac{\partial G^{\mu_{n+1}}}{\partial t}\tilde{C}_{\mu_{n+1}}(y_{n+1})}\cdot f	~.
\end{align}	
\end{lemma}
Before proving lemma \ref{Lemma2} for generic $n$, we illustrate this identity for the most simple case, namely $n=1$, which will be also relevant for linear cosmological perturbation theory. Using the Leibniz rule, we obtain
\begin{align}
\label{eq:Indn=1}
\frac{\partial G^{\mu_2}}{\partial t}X_{\tilde{C}_{\mu_1}} & X_{\tilde{C}_{\mu_2}}\cdot f
\nonumber\\
&=
X_{\tilde{C}_{\mu_1}}X_{\frac{\partial G^{\mu_2}}{\partial t}\tilde{C}_{\mu_2}}\cdot f
-X_{\tilde{C}_{\mu_1}}\tilde{C}_{\mu_2}X_{\frac{\partial G^{\mu_2}}{\partial t}}\cdot f
-\left(X_{\tilde{C}_{\mu_1}}\cdot\frac{\partial G^{\mu_2}}{\partial t}\right)X_{\tilde{C}_{\mu_2}}\cdot f,
\end{align}
where we neglected to write the arguments of the constraints and gauge fixing conditions respectively to keep our notation compact. In order to get the correct result for $n=1$, we have to show that the last two terms in the above equation vanish weakly. Let us first consider the second term on the right hand side in (\ref{eq:Indn=1}). We have
\begin{eqnarray}
-X_{\tilde{C}_{\mu_1}}\tilde{C}_{\mu_2}X_{\frac{\partial G^{\mu_2}}{\partial t}}\cdot f 
&=&
-\tilde{C}_{\mu_2}X_{\tilde{C}_{\mu_1}}X_{\frac{\partial G^{\mu_2}}{\partial t}}\cdot f
-\left(X_{\tilde{C}_{\mu_1}}\cdot\tilde{C}_{\mu_2}\right)X_{\frac{\partial G^{\mu_2}}{\partial t}}\cdot f \nonumber\\
&\approx &
-\left(X_{\tilde{C}_{\mu_1}}\cdot\tilde{C}_{\mu_2}\right)X_{\frac{\partial G^{\mu_2}}{\partial t}}\cdot f \nonumber\\
&=&
\{\tilde{C}_{\mu_1},\tilde{C}_{\mu_2}\}X_{\frac{\partial G^{\mu_2}}{\partial t}}\cdot f \nonumber\\
&\approx & 0,
\end{eqnarray}
where we used in the last line that the weakly abelianized constraints weakly commute. For the third term on the right hand side of (\ref{eq:Indn=1}), we use the fact that the clocks and secondary constraints build weakly a canonically conjugate pair. Hence, we get
\begin{align}
\label{eq:Casek=1}
-\left(X_{\tilde{C}_{\mu_1}(y_1)}\cdot\frac{\partial G^{\mu_2}}{\partial t}(y_2)\right) & X_{\tilde{C}_{\mu_2}(y_2)}\cdot f
=
-\{\frac{\partial G^{\mu_2}}{\partial t}(y_2),\tilde{C}_{\mu_1}(y_1)\}\{f,\tilde{C}_{\mu_2}(y_2)\}\nonumber \\	
&=
-\{f,\tilde{C}_{\mu_2}(y_2)\}\frac{\partial}{\partial t}\left(\{G^{\mu_2}(y_2),\tilde{C}_{\mu_1}(y_1)\}\right)\nonumber \\
&\quad\,
+\{f,\tilde{C}_{\mu_2}(y_2)\}\{G^{\mu_2}(y_2),\frac{\partial}{\partial t}\tilde{C}_{\mu_1}(y_1)\}\nonumber \\
&\approx 
\{f,\tilde{C}_{\mu_2}(y_2)\}\frac{\partial}{\partial t}\left(\delta^{\mu_2}_{\mu_1}\delta^{(3)}(y_1,y_2)\right)\nonumber \\
&\quad\,
+\int \d^3z \{f,\tilde{C}_{\mu_2}(y_2)\}\{G^{\mu_2}(y_2),{C}_{\rho}(z)\}\frac{\partial}{\partial t}{\cal B}_{\mu_1}^\rho(y_1,z) 
\nonumber\\
&=
\int \d^3z \{f,\tilde{C}_{\mu_2}(y_2)\}\{G^{\mu_2}(y_2),{C}_{\rho}(z)\}\frac{\partial}{\partial t}{\cal B}_{\mu_1}^\rho(y_1,z)\nonumber \\
&\approx  0 ~.
\end{align}
Here, we used in the third line that $\{G^\mu(y),\tilde{C}_\nu(x)\}\approx-\delta^\mu_\nu\delta^{(3)}(x,y)$ and in the last step the assumption of the lemma, namely that $\partial_t{\cal A}^\mu_\nu$ vanishes, which carries over to $\partial_t{\cal B}^\mu_\nu$ vanishing. Considering these two intermediate results, we indeed obtain for the case of $n=1$ the following weak identity:
\begin{eqnarray}
\frac{\partial G^{\mu_2}}{\partial t}X_{\tilde{C}_{\mu_1}}X_{\tilde{C}_{\mu_2}}\cdot f
&\approx &
X_{\tilde{C}_{\mu_1}}X_{\frac{\partial G^{\mu_2}}{\partial t}\tilde{C}_{\mu_2}}\cdot f .
\end{eqnarray}
Now, we consider the case of arbitrary $n$ for lemma 2. We have by means of the Leibniz rule
\begin{align}
\label{eq:Identn}
\frac{\partial G^{\mu_{n+1}}}{\partial t}(y_{n+1}) &
 X_{\tilde{C}_{\mu_n}(y_n)}\cdots X_{\tilde{C}_{\mu_1}(y_1)}X_{\tilde{C}_{\mu_{n+1}}(y_{n+1})}\cdot f &\nonumber \\
 &=
 X_{\tilde{C}_{\mu_n}(y_n)}\cdots X_{\tilde{C}_{\mu_1}(y_1)}X_{\frac{\partial G^{\mu_{n+1}}}{\partial t}\tilde{C}_{\mu_{n+1}}(y_{n+1})}\cdot f	&\nonumber \\
&\hphantom{=}\ - \left(X_{\tilde{C}_{\mu_n}(y_n)}\cdots X_{\tilde{C}_{\mu_1}(y_1)}\cdot\frac{\partial G^{\mu_{n+1}}}{\partial t}\right)
\left(X_{\tilde{C}_{\mu_{n+1}}}\cdot f\right) & \nonumber \\
&\hphantom{=}\ - X_{\tilde{C}_{\mu_n}(y_n)}\cdots X_{\tilde{C}_{\mu_1}(y_1)}\tilde{C}_{\mu_{n+1}}(y_{n+1})X_{\frac{\partial G^{\mu_{n+1}}}{\partial t}(y_{n+1})}\cdot f   ~.
 	\end{align}
Similar to the case $n=1$, we have to show that the last two terms on the right hand side of (\ref{eq:Identn}) vanish weakly. This will follow if we can show the following two results:
\begin{eqnarray}\label{eq:two-conditions}
(i) && 	X_{\tilde{C}_{\mu_k}(y_k)}\cdots X_{\tilde{C}_{\mu_1}(y_1)}\cdot \tilde{C}_{\mu_{n+1}}(y_{n+1})\approx 0\quad, \forall \, k\in\mathbf{N}_0\quad {\rm and}\nonumber \\
(ii) && X_{\tilde{C}_{\mu_k}(y_k)}\cdots X_{\tilde{C}_{\mu_1}(y_1)}\cdot\frac{\partial G^{\mu_{n+1}}}{\partial t}\approx 0,\quad \forall \, k\in\mathbf{N}.
\end{eqnarray}
Let us first consider condition $(i)$. 
For $k=0$, it just reads $\tilde{C}_{\mu_{n+1}}$, which of course weakly vanishes. Taking into account that $\{\tilde{C}_{\mu}(x),\tilde{C}_\nu(y)\}\approx 0$ as well as the first class property of the constraints $\tilde{C}_\mu(x)$, this also trivially follows for all $k\geq 1\in\mathbf{N}$. Note that the abelianized constraints $\tilde{C}_\mu$ have the property that their structure functions vanish weakly, as shown in \cite{Thiemann2}, and hence their structure functions are also a linear combination of the secondary constraint. This has the effect that the resulting terms we obtain from the multiple applications of the Hamiltonian vector fields can contain even higher than linear powers of the secondary constraints. However, for our proof, we only need that all of the terms are at least linear in the secondary constraints. 

As far as $(ii)$ is concerned, we have shown the case $k=1$ already in (\ref{eq:Casek=1}). For higher order $k > 1$, we also have to consider the terms that weakly vanish up to the point when all Hamiltonian vector fields have been applied. From the property of the clocks, we know that
\begin{eqnarray}
\{G^\mu(x),\tilde{C}_\nu(y)\}
&=&-\{T^\mu(x),\tilde{C}_\nu(y)\} \nonumber \\
&=&-\{T^\mu(x),\int \d^3z {\cal B}^\rho_\nu(y,z)C_\rho(z)\}\nonumber \\
&=&
-\int \d^3z\{T^\mu(x),C_\rho(z)\}{\cal B}^\rho_\nu(y,z)
-\int \d^3z\{T^\mu(x),{\cal B}^\rho_\nu(y,z)\}C_{\rho}(z) \nonumber \\
&=&
-\int \d^3z {\cal A}^\mu_\rho(x,z){\cal B}^\rho_\nu(y,z)
-\int \d^3z\{T^\mu(x),{\cal B}^\rho_\nu(y,z)\}C_{\rho}(z)\nonumber \\
&=&
-\delta^\mu_\nu\delta^{(3)}(x,y)
-\int \d^3z\{T^\mu(x),{\cal B}^\rho_\nu(y,z)\}C_{\rho}(z) ~.\nonumber \\
\end{eqnarray}
Therefore, we get
\begin{align}
\label{eq:InnerPB}
\{\frac{\partial G^\mu(x)}{\partial t}, & \tilde{C}_\nu(y)\}
=
\frac{\partial}{\partial t}\left(	\{G^\mu(x),\tilde{C}_\nu(y)\}\right)-\{G^\mu(x),\frac{\partial}{\partial t}\tilde{C}_\nu(y)\}\nonumber \\
&=
-\int \d^3z\frac{\partial }{\partial t}\left(\{T^\mu(x),{\cal B}^\rho_\nu(y,z)\}\right)C_{\rho}(z)
\nonumber \\
&\quad\,
+\int \d^3z \{T^\mu(x),\frac{\partial }{\partial t}{\cal B}_\nu^\rho(y,z)\}C_\rho(z)+\int \d^3z {\cal A}^\mu_\nu(x,y)\frac{\partial}{\partial t}{\cal B}_\nu^\rho(y,z)\nonumber\\
&=
-\int \d^3z\int \d^3z' \{\frac{\partial }{\partial t}T^\mu(x),{\cal B}^\rho_\nu(y,z)\}{\cal A}^\lambda_\rho(z',z)\tilde{C}_{\lambda}(z')
\nonumber\\
& \quad\, +\int \d^3z {\cal A}^\mu_\nu(x,y)\frac{\partial}{\partial t}{\cal B}_\nu^\rho(y,z) .
\end{align}
Using this for \eqref{eq:two-conditions} we can rewrite (ii) as follows:
\begin{eqnarray}
X_{\tilde{C}_{\mu_k}(y_k)}\cdots X_{\tilde{C}_{\mu_1}(y_1)}\cdot\frac{\partial G^{\mu_{n+1}}}{\partial t}(y_{n+1})
&=&
X_{\tilde{C}_{\mu_{k}}(y_{k})}\cdots X_{\tilde{C}_{\mu_{2}}(y_{2})}\cdot\{\frac{\partial G^{\mu_{n+1}}}{\partial t}(y_{n+1}),\tilde{C}_{\mu_1}(y_{1})\}	\nonumber \\
\end{eqnarray}
Considering the first term on the right hand side of (\ref{eq:InnerPB}), we realize that it is just a linear combination of the abelianized secondary constraints and hence each contribution is proportional to the secondary constraints $\tilde{C}_\mu$. Thus, we can apply $(i)$ and know that their contribution weakly vanishes. Hence, the only remaining term that could contribute is the second one on the right hand side of (\ref{eq:InnerPB}). Since, by assumption, $\partial_t{\cal A}^\mu_\nu$ vanishes weakly, there exist two cases. Either $\partial_t{\cal A}^\mu_\nu$ vanishes already strongly then we do not need to consider this term any longer. In case it vanishes only weakly, it can be expressed as some time dependent linear combination of the secondary first class constraints $\tilde{C}_\mu$ and, as a consequence of $(i)$, its contribution to $(ii)$ vanishes weakly. This finishes the proof of lemma \ref{Lemma2}. Now, we will use this lemma to prove our main result in lemma \ref{MainLemma}.

Using lemma \ref{Lemma2}, we can express the time derivative of the observable in (\ref{eq:TimeDerObsI}) as follows:
\begin{eqnarray}
\label{eq:dOdtSpec2}
\frac{\d \mathcal{O}_{f,T}}{\d t}&\approx &\sum\limits_{n=0}^{\infty} \frac{1}{n!} \int\mathrm{d}^3y_1...\int\mathrm{d}^3y_{n}~ G^{\mu_1}(y_1)...G^{\mu_n}(y_n)
\nonumber \\
&& X_{\tilde{C}_{\mu_n}(y_n)}\cdots X_{\tilde{C}_{\mu_1}(y_1)}X_{\int \d^3 y_{n+1}\frac{\partial G^{\mu_{n+1}}}{\partial t}(y_{n+1})\tilde{C}_{\mu_{n+1}}(y_{n+1})}\cdot f.
\end{eqnarray}
Next, we will show that the Hamiltonian vector field that is applied first onto $f$ in the above formula is the one associated with $H_{\rm can}$. For this purpose, we use the  result in (\ref{eq:GmuStab}) and obtain
\begin{align}
\int \d^3 y_{n+1} & \frac{\partial G^{\mu_{n+1}}}{\partial t}(y_{n+1})  \tilde{C}_{\mu_{n+1}}(y_{n+1})  \nonumber\\
& \approx \int \d^3 y_{n+1}\int \d^3z_1\int \d^3z_2 N^\rho(z_1){\cal A}^{\mu_{n+1}}_\rho(z_1,y_{n+1}){\cal B}^\lambda_{\mu_{n+1}}(z_2,y_{n+1})C_\lambda(z_2)	\nonumber \\
&=\int \d^3z_1\int \d^3z_2 N^\rho(z_1)C_\lambda(z_2)\delta^\lambda_\rho\delta^{(3)}(z_1,z_2) \nonumber\\
&=\int \d^3 z_1 N^\rho(z_1)C_\rho(z_1) \nonumber \\
&=H_{\rm can},
\end{align}
where we used in the last step that, by our assumption, $f$ does not depend on lapse and shift variables or their momenta. Let us remark that what we have used is a weak equivalence for the term $\frac{\partial G^{\mu_{n+1}}}{\partial t}$. This holds only up to terms that are linear in the first class constraint. However, since these additional terms will  be multiplied with terms also linear in the first class constraints, their contribution weakly vanishes in $\frac{\d \mathcal{O}_{f,T}}{\d t}$.

Given this, we finally obtain our main result of this section for the time derivative of the observable:
\begin{eqnarray}
\label{eq:dOdtSpecFinal}
\frac{\d \mathcal{O}_{f,T}}{\d t}&\approx &\sum\limits_{n=0}^{\infty} \frac{1}{n!} \int\mathrm{d}^3y_1...\int\mathrm{d}^3y_{n}~ G^{\mu_1}(y_1)...G^{\mu_n}(y_n)
X_{\tilde{C}_{\mu_n}(y_n)}\cdots X_{\tilde{C}_{\mu_1}(y_1)}X_{H_{\rm can}}\cdot f\nonumber \\
&=&
\sum\limits_{n=0}^{\infty} \frac{1}{n!} \int\mathrm{d}^3y_1...\int\mathrm{d}^3y_{n}~ G^{\mu_1}(y_1)...G^{\mu_n}(y_n)
X_{\tilde{C}_{\mu_n}(y_n)}\cdots X_{\tilde{C}_{\mu_1}(y_1)}\cdot \{f,H_{\rm can}\}\nonumber \\
&=&
\mathcal{O}_{\{f,H_{\rm can}\},T} ~.
\end{eqnarray}
This finishes the proof of lemma \ref{MainLemma} in the special case where the function $f$ is independent of lapse and shift degrees of freedom. If we relax this assumption, we have to repeat the above analysis for the more complicated time derivative in (\ref{eq:dOdt}). This result is proved in the appendix.

Note that in the case of gauge fixing conditions of the class in (\ref{eq:LinGF}), we have $\{G^\mu(x),C_\nu(y)\}\approx -\delta^\mu_\nu\delta^{(3)}(x,y)$. And, even for a generic $\tau^\mu$ as a function of spacetime coordinates, $\frac{\partial G^\mu}{\partial t}$ will no longer be phase space dependent because $\frac{\partial G^\mu}{\partial t}=\frac{\partial \tau^\mu}{\partial t}$, and $\tau^\mu$ is assumed to be independent of the phase space. However, in cosmological perturbation theory, already at the linear order gauge fixing conditions have a more complicated time dependence due to background FRLW quantities that are involved as pre-factors in the linearized variables. Therefore, in our case, $\frac{\partial G^\mu}{\partial t}$ will be phase space dependent. Our analysis shows that the criteria that we express the dynamics of the observables in the way shown in lemma \ref{MainLemma} is given by $\partial_t{\cal A}^\mu_\nu\approx 0$. This condition holds for all gauge fixings considered in \cite{Giesel:2018opa}, and in particular for the longitudinal and spatially gauge discussed in this work. Hence, we can apply lemma \ref{MainLemma} to derive Hamilton's equations of motion for the observables under consideration. As far as the equations of motion are considered, one of the advantages of this lemma is that we are able to derive the dynamics of the observables without explicitly knowing their Poisson algebra and the corresponding physical Hamiltonian. This is due to the following facts.  First, the algebra of the observables can be more complicated than the algebra of the gauge variant quantities. Second, in general,  it is not as straight forward to derive the physical Hamiltonian in the case of geometrical clocks as compared to models where matter reference fields have been chosen (see for instance the discussion in \cite{Dittrich-Tambornino2} in the context of the longitudinal gauge). Therefore, from this perspective, this lemma enables us to derive the dynamics of the observables in an efficient way. 
\section{Application to canonical linearized cosmological perturbation theory}
\label{Sec:Applic}
In this section, we use the result obtained in the previous section to obtain equations of motion for two specific gauges, 
 namely the longitudinal and the spatially flat gauge, which are of the form  (\ref{eq:GenGF}). In terms of the perturbations of the metric, the longitudinal gauge is identified by $E \approx 0$ and $ B \approx 0$, and the spatially flat gauge by $\psi \approx0 $ and $ E \approx 0$. Both gauges correspond to an isotropic threading of spacetime, in which the longitudinal part of the spatial metric perturbation is zero. These gauges have been recently studied in the observable formalism using geometrical clocks \cite{Giesel:2018opa}, which we refer to the reader for details on the relationships between observables and metric perturbations. In table \ref{tab:clockchoices1}, we summarize the results of \cite{Giesel:2018opa} on relationships between the metric, matter and corresponding momentum perturbations and the Dirac observables in longitudinal and spatially flat gauges. Also the corresponding geometric clocks for both gauges are stated.  Therein, in our convention, the potential $V(\bar\varphi)$ is twice the usual value of the potential, $\kappa = 16\pi G_{\mathrm{Newton}}$, $\lambda_\varphi$ is the coupling constant of the scalar field in the Hamiltonian for the minimally coupled massless scalar field and $\Ht = \mathcal{LM}(\mathcal{H}) = \frac{1}{2}\frac{\dot{A}}{A}$ is the Hubble parameter in phase space, related to $\Pt$ via $\Ht = -\frac{2\sqrt{A}}{\Nb \Pt}$.

The equations of motion for the matter degrees of freedom then read
\begin{align}
\dot{\bar\varphi} &= \lphi\frac{\Nb}{A^{\nicefrac{3}{2}}}\piphibar & {\rm and} & & \dot{\bar\pi}_\varphi &= -\frac{\Nb A^{\nicefrac{3}{2}}}{2\lphi}V'(\bar\varphi) . \label{eq:EOMmatter}
\end{align}
We also introduce $\rho$ and $p$ as the energy density and pressure of the scalar field,
\begin{align}
\rho &= \frac{1}{2}\left( \frac{\lambda_\varphi}{A^3}\bar{\pi}_\varphi^2 + \frac{1}{\lambda_\varphi}V(\bar{\varphi}) \right) & {\rm and}& & p &=\frac{1}{2}\left( \frac{\lambda_\varphi}{A^3}\bar{\pi}_\varphi^2 - \frac{1}{\lambda_\varphi}V(\bar{\varphi}) \right), \label{eq:prho}
\end{align}
where $\rho$ and $p$ satisfy the Friedmann equations
\begin{align}
\Ht^2 &= \frac{\kappa}{6} \bar N^2 \rho  & {\rm and} & & \dot \Ht = \frac{\dot{\bar N}}{\bar N} \Ht - \frac{3}{2} \Ht^2 - \frac{\kappa}{4} \bar N^2 p ~. 
 \label{eq:Friedmann}
\end{align}

\begin{table}[tbh!]
\begin{center}
\renewcommand{\arraystretch}{1.5}
\begin{tabular}{c||c|c}
Variable & Longitudinal & Spatially flat \\
\hline \hline
$\delta T^0$ & $2\tilde{P}\sqrt{A}(E+p_E)$ & $\frac{\bar{N}}{\tilde{\mathcal{H}}}(\psi-\frac{1}{3}\Delta E)$ \\
$\delta T^a$ & $\delta^{ab}(E_{,b}+F_b)$ & $\delta^{ab}(E_{,b}+F_b)$  \\
\hline
$\phi$ & $-\Psi$ &  $- 2\Upsilon - \left(\frac{1}{2} + \frac{\kappa}{\tilde P^2} A p\right) \Psi $   \\
$B$ & $0$ & $\frac{\bar{N}^2}{A\tilde{\mathcal{H}}}\Psi$ \\
$S^a$ & $4\tilde{\mathcal{H}}\nu^a$ & $4\tilde{\mathcal{H}}\nu^a$ \\
$\psi$ & $\Psi$ & $0$ \\
$E$ & $0$ & $0$  \\
$F_a$ & $0$ & $0$ \\
$p_\psi$ & $\Upsilon$ & $\Upsilon + \alpha \Psi$ \\
$p_E$ & $0$ & $\frac{1}{\tilde{P}^2}\Psi$ \\
$p^a_F$ & $\nu^a$ & $\nu^a$ \\
$\delta \varphi$ & $\delta\varphi^{(gi)}$ & $v$ \\
$\delta \pi_\varphi$ & $\delta\pi_\varphi^{(gi)}$ & $\pi_v$ \\
$p_\phi$ & $\frac{1}{\Nb}\delta\Pi$ & $\frac{1}{\Nb}\delta\Pi$ \\
$p_B$ & $\delta\hat{\Pi}$ & $\delta\hat{\Pi}$ \\
$p_S^a$ & $\delta\Pi^a_\perp$ & $\delta\Pi^a_\perp$ \\
\end{tabular}
\caption{This table summarizes results of geometrical clocks and linearized observables corresponding to various metric perturbations and their momenta for the longitudinal and the spatially flat gauges obtained in \cite{Giesel:2018opa}. Symbols used are summarized in table \ref{tab:symbols}. \label{tab:clockchoices1}
} 
\end{center}
\end{table}
\begin{table}[h!]
 \begin{center}
  \renewcommand{\arraystretch}{1.5}
\begin{tabular}{c||c}
Symbol & Relation to background and perturbation variables  \\
\hline \hline
$\alpha$ & $\frac{1}{4} + \frac{\kappa}{2 \tilde P^2} A p -\frac{2}{3}\frac{1}{\tilde{P}^2}\Delta$  \\
$\nu^a$ & $p_F^a(x)+\delta^{ab}F_b(x)$ \\
$\Psi$ &    $\psi(x)+\frac{4\tilde{\mathcal{H}}^2A}{\bar{N}^2}(E+p_E)(x)-\frac{1}{3}\Delta E(x)  $    \\
$\Upsilon$ & $ p_\psi +\frac{\Delta E}{2} + \frac{2}{3}\Delta p_E - \left( \frac{\tilde{\mathcal{H}}^2A}{\bar{N}^2}+\frac{\kappa A}{2} p \right)(E+p_E)$ \\
$v$ & $\delta \varphi -\frac{\lambda_\varphi}{A^{\nicefrac{3}{2}}\tilde{\mathcal{H}}}\bar{N}\bar{\pi}_\varphi \left( \psi -\frac{1}{3}\Delta E\right)$ \\
$\pi_v$ & $\delta \pi_\varphi - \bar{\pi}_\varphi\Delta E + \frac{1}{2}\frac{A^{\nicefrac{3}{2}}}{\lambda_\varphi} V'(\bar\varphi) \frac{\bar{N}}{\tilde{\mathcal{H}}}\left( \psi -\frac{1}{3}\Delta E \right)$ \\
\end{tabular}
\caption{Various symbols used in table \ref{tab:clockchoices1} defined in terms of background and perturbed quantities, where $V'(\bar\varphi) \coloneqq \frac{\mathrm{d}V}{\mathrm{d}\varphi}(\bar{\varphi})$}  
\label{tab:symbols}
\end{center}
 \end{table}

The equations of motion for various scalar and vector components of metric perturbations and of their momenta can be systematically found using the Hamiltonian formulation~\cite{Giesel:2018opa}. For the scalar components of the metric perturbations, these 
are given by
\begin{equation}
\label{eq:dotpsi}
\dot{\psi} = 2\tilde{\mathcal{H}}\left( p_\psi - \frac{1}{2}\psi \right) + \tilde{\mathcal{H}}\phi + \frac{1}{3}\Delta B
\end{equation}
and
\begin{equation}\label{eq:dotE}
\dot{E} = -4\tilde{\mathcal{H}}(E+p_E) + B.
\end{equation}
Time derivatives of their corresponding momenta are
\begin{eqnarray}
\label{eq:dotppsi}
\dot{p}_\psi &=& \nonumber \frac{1}{6}\frac{\bar{N}^2}{A\tilde{\mathcal{H}}}\Delta\left( \phi+\psi-\frac{1}{3}\Delta E \right) +\left( -\frac{1}{2}\tilde{\mathcal{H}} + \frac{\kappa}{4}\frac{\bar{N}^2}{\tilde{\mathcal{H}}}p \right)\left( p_\psi-\frac{1}{2}\psi \right) - \frac{\kappa}{8}\frac{\bar{N}^2}{\tilde{\mathcal{H}}}\delta\tilde{T} \\
&& -\frac{1}{2}\left( \frac{1}{2}\tilde{\mathcal{H}} + \frac{\kappa}{4}\frac{\bar{N}^2}{\tilde{\mathcal{H}}}p \right)\phi + \frac{1}{6}\Delta B, 
\end{eqnarray}
and
\begin{equation}\label{eq:dotpE}
\dot{p}_E = -\frac{1}{4}\frac{\bar{N}^2}{A\tilde{\mathcal{H}}} \left( \phi+\psi-\frac{1}{3}\Delta E \right)+\left(\frac{5}{2}\tilde{\mathcal{H}} + \frac{\kappa}{4}\frac{\bar{N}^2}{\tilde{\mathcal{H}}}p \right) (E+p_E) -B~.
\end{equation}
Here, $\delta \tilde{T}$ denotes the spatial energy momentum perturbation\footnote{We want to point out that in~\cite{Giesel:2017roz} in equation (3.97) the prefactor $\frac{1}{A^{\nicefrac{3}{2}}}$ is a typo. However, in what follows in~\cite{Giesel:2017roz}, the correct version of $\delta\tilde{T}$ was used e.g. given in (C7) therein.}
\begin{equation}
\delta \tilde{T} \coloneqq  -3\lphi\frac{\piphibar^2}{A^3}\psi + \lphi\frac{\piphibar}{A^3}\delta\pi_\varphi - \frac{1}{2\lphi}V'(\bar\varphi)\delta\varphi . \label{eq:deltaT}
\end{equation}

The equations of motion for the vector component of the metric perturbation and its momentum are
\begin{equation}\label{eq:dotFpF}
\frac{\mathrm{d}}{\mathrm{d}t} \left[ \begin{matrix}
\delta^{ab}F_b \\
p_F^a
\end{matrix} \right] = \left[ \begin{matrix}
-4\tilde{\mathcal{H}} & -4\tilde{\mathcal{H}} \\
\digamma & \digamma
\end{matrix} \right] \left[ \begin{matrix}
\delta^{ab}F_b \\
p_F^a
\end{matrix} \right] + \left[ \begin{matrix}
S^a \\
-S^a
\end{matrix} \right] ,
\end{equation}
where $\digamma \coloneqq  (10 \tilde {\cal H}^2 + \kappa \bar N^2 p)/4 \tilde{\cal H}$. 

The time derivatives of the perturbed lapse and shift perturbations are given by
 \begin{align}
\label{eq:dotphidotB}
\dot{\phi} &= -\frac{\dot{\bar{N}}}{\bar{N}}\phi + \frac{\delta \lambda}{\bar{N}} \ , & \dot{B} &= \delta \hat{\lambda} & {\rm and} && \dot{S}^a &= \delta \lambda^a_\perp ,
\end{align}
whereas the corresponding momenta of lapse and shift satisfy
\begin{align}\label{eq:dotpphi}
\dot{p}_\phi & = \tdif\left(\frac{1}{\bar N} \delta \Pi \right)  = - \frac{1}{\bar N} (\dot {\bar N} p_\phi + \delta C) \ , \\
\dot p_B &= \delta \dot {\hat \Pi} =  -\delta {\hat C} \qquad  {\rm and} \qquad \dot p_{S^a} =  \delta \dot \Pi_\perp^a  = - \delta C_a^\perp ~. \label{eq:dotpB}
 \end{align}
Further, the perturbed Hamiltonian and spatial diffeomorphism constraints for the perturbed FLRW metric turn out to be 
\begin{align}
\label{eq:deltaCdeltaCa}
\delta C &=  -\frac{3}{2}\sqrt{A}\tilde{P}^2(\psi+4 \,p_\psi) + 4\sqrt{A}\Delta\left( \psi -\frac{1}{3}\Delta E \right) \nonumber \\ 
&~~~~\!+ \kappa A^{3/2} \left( -3 \, p \, \psi +  \frac{\lambda_\varphi}{A^3}\bar{\pi}_\varphi \delta \pi_\varphi + \frac{1}{2\lambda_\varphi}V'(\bar\varphi)\delta \varphi \right), \nonumber \\
\delta \hat{C} &= -4A\tilde{P}\left( p_\psi - \frac{1}{2}\psi + \frac{2}{3}\Delta (E+p_E) \right) + \kappa \bar{\pi}_\varphi\delta \varphi \ \ {\rm and} \nonumber \\
\delta C^\perp_a &= 2A\tilde{P}(F_a+p_F^b\delta_{ab}).
\end{align}

Finally, we will need the equations of motion for the perturbation of the scalar field and its momentum\footnote{We want point out that in~\cite{Giesel:2017roz} in equation (3.98) and in \cite{Giesel:2018opa} in equation (2.46) the last term involving $\delta B$ is missing. The results in \cite{Giesel:2018opa, Giesel:2017roz} however involve this contribution, because either equation (3.68) in \cite{Giesel:2018opa} was used where this term is taken into account or the second equation was only used for the longitudinal gauge where this term vanishes anyway.}:
\begin{align}\label{eq:dotvarphi}
\delta \dot{\varphi} &= \bar{N}\frac{\lambda_\varphi}{A^{3/2}}\bar{\pi}_\varphi\left( \phi -3\psi + \frac{\delta \pi_\varphi}{\bar{\pi}_\varphi} \right) \ \ {\rm and} \nonumber \\
\delta \dot{\pi}_\varphi &= \bar{N}\frac{A^{3/2}}{\lambda_\varphi}\left[  -\frac{1}{2}V'(\bar\varphi)(\phi+3\psi) +\frac{1}{A}\Delta\delta\varphi -\frac{1}{2}V''(\bar\varphi)\delta\varphi \right] + \piphibar\Delta B.
\end{align}

\subsection{Dynamical equations for the longitudinal gauge}
For the longitudinal gauge, the perturbed clocks are \cite{Giesel:2018opa}
\begin{equation}
\delta T^0 = 2\tilde{P}\sqrt{A}(E+p_E) \qquad {\rm and}\qquad \delta T^a = \delta^{ab}(E_{,b}+F_b) ~.
\end{equation}
It is straightforward to see that the conditions $\delta T^0 \approx 0$ and $\delta \hat T \approx 0$ along with their temporal stability conditions are equivalent to the longitudinal gauge conditions and conditions for its stability on the phase space: $E \approx 0$, $B \approx 0$, $p_E \approx 0$ and $\psi \approx - \phi$.  In the following, we use the observables found using above geometrical clocks to obtain their Hamilton's equations of motion. This is then followed by deriving the corresponding second order equation for the Bardeen potential.

\subsubsection{Hamilton's Equations for the longitudinal gauge}
In this subsection, we derive the gauge invariant Hamilton's equations of motion for all observables in the longitudinal gauge. The procedure uses various equations for the gauge variant perturbations in the lapse, shift and spatial metric degrees of freedom and their momenta, along with their time derivatives which have been obtained in \cite{Giesel:2018opa}. A crucial property of observables that we will often use in the further computations is (cf.  e.g. \cite{Giesel:2012rb})
\be\label{Oprop1}
\mathcal{O}_{F(q,p),T} \approx F(\mathcal{O}_{q,T}, \mathcal{O}_{p,T}) ~. 
\ee
Let us start with the observable for the trace part of the spatial metric perturbation, denoted by $\psi$. Using lemma \ref{MainLemma}, we can then find the time derivative of $\Obs{\psi}$ using  \eqref{eq:dotpsi} as follows:
\begin{eqnarray}\label{dotpsi}
\tdif\Obs{\psi} &\approx& \nonumber \Obs{\{\psi,H_{\rm ADM}\}} = \Obs{\dot \psi} \\
&=& \nonumber \Obs{2\tilde{\mathcal{H}}\left( p_\psi - \frac{1}{2}\psi \right) + \tilde{\mathcal{H}}\phi + \frac{1}{3}\Delta B} \\
&\approx& \nonumber 2 \Ht \Obs{p_\psi} - \Ht \Obs{\psi} + \Ht \Obs{\phi} \\
&=&  2\Ht\left( \Upsilon- \Psi \right) ~.
\end{eqnarray}
Here, we have used (\ref{Oprop1}), the vanishing of $\mathcal{O}_{\frac{1}{3}\Delta B, T}$ in the longitudinal gauge in the third step and the observable formulae for gauge invariant variables in table \ref{tab:clockchoices1} in the last equation.

The equation of motion for the observable corresponding to longitudinal scalar part of the spatial metric perturbation turns out to be 
\begin{eqnarray}\label{doE}
 \tdif\Obs{E} &\approx& \nonumber \Obs{\{E,H_{\rm ADM}\}} = \Obs{\dot E} \\
 &=& - 4 \Ht  (\Obs{E} + \Obs{p_E}) + \Obs{B} \approx 0, 
\end{eqnarray}
using results from table \ref{tab:clockchoices1}. The last equations shows that, as expected, the stability of the gauge fixing condition is also implemented at the level of observables.

The equations for the time derivative of the observables corresponding to the momentum of $\psi$ and $E$ can be derived similarly. For $\Obs{p_\psi}$, we get
\begin{eqnarray}\label{dotppsi}
\tdif\Obs{p_\psi} &\approx&  \nonumber \Obs{\{p_\psi,H_{\rm ADM}\}} \\
&=& \nonumber \frac{1}{6} \frac{\bar N^2}{A \Ht} (\Obs{\Delta \phi} + \Obs{\Delta \psi} - \Obs{\frac{\Delta^2 E}{3}}) + \left(\frac{\kappa}{4} \frac{\bar N^2}{\Ht} p - \frac{\Ht}{2} \right) (\Obs{p_\psi} - \Obs{\frac{\Psi}{2}}) \\
&& \nonumber -\frac{1}{2} \left(\frac{\kappa}{4} \frac{\bar N^2}{\Ht} p + \frac{\Ht}{2} \right) \Obs{\phi} + \frac{1}{6} \Obs{\Delta B} - \frac{\kappa}{8} \frac{\bar N^2}{\Ht} \Obs{\delta \tilde T} \\
&\approx&  \left( -\frac{\Ht}{2} + \frac{\kappa}{4} \frac{\Nb^2p}{\Ht} \right)\Upsilon + \frac{\Ht}{2}\Psi - \frac{\kappa}{8}\frac{\Nb^2}{\Ht}\dT \ ,
 \end{eqnarray}
where $\dT \coloneqq  \Obs{\delta \tilde T}$. Above, in the second step, we have used  (\ref{eq:dotppsi}) and in the final step the longitudinal gauge condition, which implies $\Obs{\psi} \approx - \Obs{\phi}$, along with relations between observables and gauge invariant variables. The equation of motion for $\Obs{p_E}$ becomes with~(\ref{eq:dotpE})
\begin{eqnarray}
\tdif\Obs{p_E} &\approx& \nonumber \Obs{\{p_E,H_{\rm ADM}\}} \\
&=& \nonumber - \frac{1}{4} \frac{\bar N^2}{A \Ht} (\Obs{\phi} + \Obs{\psi} - \Obs{\frac{\Delta E}{3}}) + \left(\frac{5}{2}\Ht + \frac{\kappa}{4} \frac{\bar N^2}{\Ht} p \right) (\Obs{E} + \Obs{p_E}) - \Obs{B} \\
&\approx& 0,
\end{eqnarray}
on using conditions for the longitudinal gauge. 

For the vector component of the metric perturbation and its momentum, using  (\ref{eq:dotFpF}) and relations in table \ref{tab:clockchoices1}, we get
\begin{eqnarray}
\tdif\Obs{F^a} &\approx& \nonumber \Obs{\{F^a,H_{\rm ADM}\}} 
=  - 4 \Ht(\Obs{F^a} + \Obs{p_F^a}) + \Obs{S^a} \\
&=& 0
\end{eqnarray}
and 
\begin{eqnarray}\label{eq:dotOpf}
\tdif\Obs{p_F^a} &\approx& \nonumber \Obs{\{p_F^a,H_{\rm ADM}\}} 
 = \digamma(\Obs{F^a} + \Obs{p_F^a}) - \Obs{S^a} \\
&=& (\digamma - 4 \Ht) \nu^a ~.
\end{eqnarray}

In case of perturbations of lapse and shift, after using  (\ref{eq:dotphidotB}), we obtain
\begin{eqnarray}\label{eq:dotOphi}
\tdif\Obs{\phi} &\approx& \nonumber \Obs{\{\phi,H_{\rm ADM}\}} 
= -\frac{\dot{\bar N}}{\bar N} \Obs{\phi} + \frac{1}{\bar N}\Obs{\delta \lambda} \\
&=& \frac{\dot{\bar N}}{\bar N} \Psi + \frac{1}{\bar N}\Obs{\delta \lambda} ~,
\end{eqnarray}
\begin{eqnarray}
\tdif\Obs{B} \approx \Obs{\{B,H_{\rm ADM}\}} = \Obs{\delta \hat \lambda} = 0,
\end{eqnarray}
and 
\begin{eqnarray}
\tdif\Obs{S^a} \approx \Obs{\{S^a,H_{\rm ADM}\}} = \Obs{\delta \lambda^a_\perp} .
\end{eqnarray}
The observables corresponding to Lagrange multipliers can be determined as follows. The simplest of these is $\Obs{\delta \hat\lambda}$, which vanishes owing to the equation of motion for $\Obs{B}$. The same is not true for $\Obs{\delta \lambda}$ and $\Obs{\delta \lambda^a_\perp}$. Let us note that
$\Obs{\phi} \approx - \Obs{\psi}$, which implies $\tdif\Obs{\phi} \approx -\tdif\Obs{\psi}$. Using this together with the observable equivalent of~(\ref{eq:dotphidotB}),
\begin{align}
 \tdif\Obs{\phi} & \approx -\frac{\dot{\Nb}}{\Nb}\Obs{\phi} + \frac{1}{\Nb}\Obs{\dl}~, \label{dotOphi}
 \end{align}
and~(\ref{dotpsi}), we get
\be \Obs{\dl} = \left(2\Ht\Nb-\dot{\Nb}\right)\Psi - 2\Ht\Nb\Upsilon \ . \ee 
Thus, we can rewrite~(\ref{dotOphi}) as 
\be 
\tdif\Obs{\phi} = 2 \Ht(\Psi - \Upsilon) ~.
\ee
Similarly, noting that $\Obs{S^a} = 4\Ht \Obs{{p_F}^a}$ implies $\tdif\Obs{S^a} = 4\Ht \tdif \Obs{{p_F}^a} - \frac{\dot\Ht}{\Ht} \nu^a$, we get
\be \Obs{\dl^a_\perp} = \left( 4\dot{\Ht} - 16\Ht^2+4\digamma\Ht \right)\nu^a \ , \ee
which equals the time derivative of $\Obs{S^a}$. 

The equation of motion for the momentum of the perturbation in the lapse function turns out to be 
\begin{eqnarray}
\tdif\Obs{p_\phi} &\approx& \nonumber  \Obs{\{p_\phi,H_{\rm ADM}\}} = -\frac{\dot{\bar N}}{\bar N^2} \Obs{\delta \Pi} - \frac{1}{\bar N} \delta C \\
&\approx& - \frac{\dot{\bar N}}{\bar N^2} \delta \Pi\approx 0 \ ,
\end{eqnarray}
where we have used  (\ref{eq:dotpphi}) and the vanishing of $\delta C$ and $\delta \Pi$. Similarly, time derivatives of the momenta of the components of the shift perturbation, using  (\ref{eq:dotpB}), become
\be
\tdif\Obs{p_B} \approx \Obs{\{p_B,H_{\rm ADM}\}} = \Obs{\delta{\hat\Pi}} = \delta \hat \Pi \approx 0
\ee
and
\be
\tdif\Obs{{p_S}_a} \approx \Obs{\{S^a,H_{\rm ADM}\}} = \Obs{\delta \Pi^a_\perp} = \delta \Pi^a_\perp \approx 0.
\ee
Finally, the equations of motion for the scalar field and its momentum perturbation can be found using  (\ref{eq:dotvarphi}). These turn out to be
\begin{eqnarray}
\tdif\Obs{\delta \varphi} &\approx& \nonumber \Obs{\{\delta\varphi,H_{\rm ADM}\}} = \bar N \frac{\lambda_\varphi}{A^{3/2}} \bar \pi_\varphi \left(\Obs{\phi} - 3 \Obs{\psi} + \frac{1}{\bar \pi_\varphi} \Obs{\delta \pi_\varphi}\right) \\
&=&  \bar N \frac{\lambda_\varphi}{A^{3/2}} \bar \pi_\varphi\left(- 4 \Psi + \frac{\delta \pi_\varphi^{(\mathrm{gi})}}{\bar \pi_\varphi}\right)
\end{eqnarray}
and 
\begin{align}
\tdif\Obs{\delta\pi_\varphi} &\approx \nonumber \Obs{\{\delta\pi_\varphi,H_{\rm ADM}\}}\\
& = -\bar N \frac{\lambda_\varphi}{A^{3/2}} \left(\frac{1}{2} V'(\bar\varphi)  (\Obs{\phi} + \Obs{3 \psi}) - \frac{1}{A} \Obs{\Delta \delta \varphi} + \frac{1}{2}  V''(\bar\varphi) \Obs{\delta\varphi}\right) + \piphibar\Delta\Obs{B} \nonumber\\ 
& \approx -\bar N \frac{\lambda_\varphi}{A^{3/2}} \left(V'(\bar\varphi) \Psi - \frac{1}{A} \Delta \delta \varphi^{(\mathrm{gi})} + \frac{1}{2}  V''(\bar\varphi) \delta \varphi^{(\mathrm{gi})} \right).
\end{align}

\subsubsection{Second order equation for the Bardeen potential}
We now illustrate the way our formalism yields directly the evolution equation for the Bardeen potential. To derive this equation, we note that
\be\label{ddotpsi}
\ddot \Psi = \tdif \left(\tdif \Obs{\psi} \right) \approx - 2 \tdif (\Ht (\Psi - \Upsilon)) ,
\ee
where $\Upsilon$ is the gauge invariant observable corresponding to $p_\psi$ (see table \ref{tab:clockchoices1}). 
Using the Friedmann equations \eqref{eq:Friedmann}  together with  (\ref{dotppsi}) in  (\ref{ddotpsi}), we get
\begin{eqnarray}
\ddot \Psi &\approx& \nonumber   -2 \Ht \dot \Psi - \left(2 \frac{\dot{\bar N}}{\bar N} \Ht - 3 \Ht^2 - \frac{\kappa}{2} \bar N^2 p\right) (\Psi - \Upsilon) \\
&& + \Ht^2 \Psi - \frac{\kappa}{4} \bar N^2 \dT + 2 \Ht \left(-\frac{\Ht}{2} + \frac{\kappa}{4} \frac{\bar N^2 p}{\Ht} \right)\Upsilon ~. 
\end{eqnarray}
This quickly simplifies to
\be
\ddot \Psi \approx \left(\frac{\dot{\bar N}}{\bar N} - 4 \Ht \right) \dot \Psi + \frac{\kappa}{2} \bar N^2 p \Psi - \frac{\kappa}{4} \bar N^2 \dT ~, \label{eq:StartBardeen}
\ee
where we now need to find an expression for $\dT$ in terms of $\Psi$ only. Since this requires some lengthy but straightforward steps, we refer to Appendix~\ref{deriv:Bardeen} for details and just state the final form for the second order equation of motion for the Bardeen potential:
\begin{eqnarray}
    \ddot\Psi  \approx \frac{\Nb^2}{A}\Delta\Psi + \left( \frac{\kappa}{2}\Nb^2 p - \frac{\Nb^2}{\dot{\bar\varphi}} V'(\bar\varphi) - 3\Ht^2 \right)\Psi    + \left( \frac{\dot\Nb}{\Nb} - 7\Ht - \frac{\Nb^2\Ht}{\dot{\bar\varphi}} V'(\bar\varphi) \right)\dot\Psi . \nonumber\\
    \label{eq:Bardeen}
\end{eqnarray}

We thus see our formalism yields the Lagrangian evolution equation in a simple and straightforward way when compared to the conventional derivation.

While (\ref{eq:StartBardeen}) is in correspondence with the result of~\cite{Mukhanov} (cf. the third equation of (4.15) therein, with $D=0$, $k=0$), our final second order equation equals the one in~\cite{Brandenberger} (cf. (62) therein)\footnote{Note that in ~\cite{Brandenberger} they use conformal time $\Nb=\sqrt{A}$, implying $\Ht = \frac{\dot{\Nb}}{\Nb}$, $\dot{\Ht}=-\frac{\Ht}{2}-\frac{\kappa}{4}Ap$ and $\ddot{\bar\varphi}=-2\Ht\dot{\bar\varphi}-\frac{A}{2}V'(\bar\varphi)$.}.

Lastly, we note that $\Psi$ is one of the three remaining independent degrees of freedom in the longitudinal gauge, with the other two being covered by the perturbed metric's tensorial part, $\Obs{h^{TT}_{ab}}$.

\subsection{Dynamical equations for the spatially flat gauge}
We now perform the same steps for the spatially flat gauge defined by
\begin{equation}
    \psi \approx 0 \qquad \text{and} \qquad E \approx 0 ,
\end{equation}
where we will end up with a second order equation for the Mukhanov-Sasaki variable.
The perturbed clocks read in this gauge~\cite{Giesel:2018opa}
\begin{equation}
    \delta T^0 = \frac{\Nb}{\Ht}\left( \psi - \frac{1}{3}\Delta E \right) \qquad \text{and} \qquad \delta T^a = \delta^{ab}(E_{,b}+F_b) .
\end{equation}
Demanding $\delta T^0\approx 0$ and $\delta {\hat T} \approx 0$ together with the corresponding stability conditions $\delta \dot{T}^0\approx 0$ and $\delta \dot{\hat T} \approx 0$ result in the above stated gauge conditions $\psi \approx 0$ and $E \approx 0$ as well as
\begin{equation}
    \phi \approx -2p_\psi - \frac{4}{3}\Delta p_E \qquad \text{and} \qquad B \approx 4\Ht p_E . \label{eq:stabilitySFG}
\end{equation}
The next subsection introduces the Hamilton's equations for all observables in table~\ref{tab:clockchoices1}, followed by the derivation of the Mukhanov-Sasaki equation in the subsequent subsection.

\subsubsection{Hamilton's Equations for the spatially flat gauge}
In this subsection, we derive the gauge invariant Hamilton's equations of motion for all observables in the spatially flat gauge. The procedure is the same as for the longitudinal gauge -- the time derivative is firstly moved to the perturbation the observable map acts on, then the equation of motion thereof is inserted and finally we apply the observable map according to table \ref{tab:clockchoices1}, as depicted here for $\Obs{\phi}$:
\begin{align}
\tdif \Obs{\phi}  & \approx \Obs{\dot\phi} = \Obs{-\frac{\dot\Nb}{\Nb}\phi+\frac{\dl}{\Nb}} \nonumber \\
  &= -\frac{\dot\Nb}{\Nb}\left( -2\Upsilon-\left(\frac{1}{2}+\frac{\kappa}{\Pt^2}pA\right)\Psi \right) + \frac{1}{\Nb}\Obs{\delta\lambda} . \nonumber
\end{align}
By the same method, we obtain the full set
\begin{align}
&\tdif\Obs{\phi} \approx -\frac{\dot\Nb}{\Nb}\left( -2\Upsilon-(\frac{1}{2}+\frac{\kappa}{\Pt^2}pA)\Psi \right) + \frac{1}{\Nb}\Obs{\delta\lambda} \nonumber  \\
& \hphantom{\tdif\Obs{\phi}} \overset{(\ref{SFGdl})}{\approx} -\frac{\kappa}{2}\frac{\Nb^2p}{\Ht}\left( 2\Upsilon + \left( \frac{1}{2}+\frac{\kappa}{4}\frac{\Nb^2p}{\Ht^2} \right)\Psi \right) + \frac{\kappa}{4}\frac{\Nb^2}{\Ht}\dT \ , \\
&\tdif\Obs{p_\phi} \approx 0  \ ,  \\
&\tdif\Obs{\psi} \approx 0 \ ,  \\
& \tdif\Obs{p_\psi} \approx \frac{1}{6}\frac{\Nb^2}{A\Ht}\Delta\Psi + \left( \frac{\kappa}{8}\frac{\Nb^2p}{\Ht}+\frac{\kappa^2}{16}\frac{\Nb^4p^2}{\Ht^3} \right)\Psi  \nonumber\\
&\hphantom{\tdif\Obs{p_\psi} =} -\frac{1}{3}\frac{\Nb^2}{A\Ht}\Delta\Upsilon + \frac{\kappa}{2}\frac{\Nb^2p}{\Ht}\Upsilon - \frac{\kappa}{8}\frac{\Nb^2}{\Ht}\dT \ ,  \\  
&\tdif\Obs{E} \approx 0 \ ,  \\
& \tdif\Obs{p_E} \approx \frac{1}{2}\frac{\Nb^2}{A\Ht}\Upsilon + \left( \frac{\kappa}{8}\frac{\Nb^4p}{A\Ht^3} - \frac{1}{4}\frac{\Nb^2}{A\Ht}\right)\Psi  \ ,  \\
&\tdif\Obs{B} \approx \Obs{\dlh} \overset{(\ref{SFGdlh})}{\approx} \left( \frac{\dot{\Nb}\Nb}{A\Ht} - 3\frac{\Nb^2}{A} \right) \Psi + \frac{\Nb^2}{A}\left( 2\Upsilon + \left( \frac{1}{2}+\frac{\kappa}{4}\frac{\Nb^2p}{\Ht^2} \right)\Psi \right)\ ,  \\
& \tdif\Obs{p_B} \approx 0  \nonumber  \\
\end{align}
and for the scalar field degrees of freedom
\begin{align}
&\tdif\Obs{\dphi} \approx 
\lphi\frac{\Nb\piphibar}{A^{\frac{3}{2}}}\left( -2\Upsilon - \left( \frac{1}{2}+\frac{\kappa}{4}\frac{\Nb^2 p }{\Ht^2} \right) \Psi+  \frac{\piv}{\piphibar} \right)  \quad \text{and}  \\
& \tdif\Obs{\dpiphi} \approx
\frac{\Nb A^{\frac{3}{2}}}{\lphi} \left( \frac{1}{A}\Delta v - \frac{1}{2}V''(\bar\varphi)v + V'(\bar\varphi)\Upsilon\right) \nonumber \\
& \hphantom{\tdif\Obs{\dpiphi}=} + \frac{\Nb A^{\frac{3}{2}}}{\lphi}V'(\bar\varphi) \left( \frac{1}{4} +\frac{\kappa}{8}\frac{\Nb^2p}{\Ht^2} \right)\Psi + \piphibar \frac{\Nb^2}{A\Ht}\Delta\Psi ,
\end{align}
where we already inserted later results for $\Obs{\delta\lambda}$,~(\ref{SFGdl}), and $\Obs{\dlh}$,~(\ref{SFGdlh}), to make the list complete at this stage.

The vector sector of the spatially flat gauge coincides with the one of the longitudinal gauge:
\begin{align}
\tdif\Obs{S^a} &\approx \Obs{\dl^a_{\perp}} \ , &\tdif\Obs{{p_S}_a} &\approx 0  \ ,  \\
\tdif\Obs{F_a} &\approx 0 \qquad\qquad \text{and}  &\tdif\Obs{{p_F}^a} &\approx \left(\digamma-4\Ht\right)\nu^a \ .
\end{align}

Now, we want to determine the observables for the Lagrange multipliers $\dl$, $\dlh$ and $\dl^a_\perp$. Since the calculations are  a bit more elaborate than in the longitudinal gauge, we will show the essential steps.
Starting with $\Obs{\dl}$, we first recap that this observable entered through the equation of motion of $\Obs{\phi}$,
\begin{align}
 \tdif\Obs{\phi} & \approx -\frac{\dot{\Nb}}{\Nb}\Obs{\phi} + \frac{1}{\Nb}\Obs{\dl} \ . \tag{\ref{dotOphi}}
 \end{align}
We then notice that we can reformulate
\begin{align}
\Obs{\phi} & = -2\left[ \Upsilon + \left( \frac{1}{4} + \frac{\kappa}{2\Pt^2}pA \right)\Psi \right] \nonumber\\
 \hphantom{\Obs{\phi}} & = -2\left[ \Upsilon + \left( \frac{1}{4} + \frac{\kappa}{2\Pt^2}pA \right)\Psi -\frac{2}{3} \frac{1}{\Pt^2}\Delta\Psi \right] - \frac{4}{3}\frac{1}{\Pt	^2}\Delta\Psi \nonumber\\
 & = -2\Obs{p_\psi} - \frac{4}{3}\Delta\Obs{p_E} \  \label{OphiSFG}
\end{align}
and, with that, solve~(\ref{dotOphi}) for $\Obs{\dl}$ and replace $\Obs{\phi}$ according to~(\ref{OphiSFG}) as well as the respective expressions of the time derivatives of $\Obs{p_\psi} $ and $\Obs{p_E}$ to get
\begin{align}
\Obs{\dl} &\approx \dot{\Nb}\Obs{\phi} + \Nb \tdif\Obs{\phi} = \dot{\Nb}\Obs{\phi} + \Nb \tdif \left( -2\Obs{p_\psi} - \frac{4}{3}\Delta\Obs{p_E} \right)  \nonumber\\
&= \left( \dot{\Nb}+\frac{\kappa}{2}\frac{\Nb^3p}{\Ht} \right)\Obs{\phi} + \frac{\kappa}{4}\frac{\Nb^3}{\Ht}\dT \\
&= \left( \dot{\Nb}+\frac{\kappa}{2}\frac{\Nb^3p}{\Ht} \right)\left( -2\Upsilon - \left( \frac{1}{2}+\frac{\kappa}{4}\frac{\Nb^2p}{\Ht^2} \right)\Psi \right) + \frac{\kappa}{4}\frac{\Nb^3}{\Ht}\dT \label{SFGdl} \ .
\end{align}

Proceeding with $\Obs{\dlh}$, we first notice
\be \tdif\Obs{p_E} \approx -\frac{\Nb^2}{4A\Ht}\Obs{\phi}-\left( \frac{3}{2}\Ht - \frac{\kappa}{4}\frac{\Nb^2p}{\Ht} \right)\Obs{p_E} \ . \ee
Hence, using the expression of $ \dot\Ht $,~\eqref{eq:Friedmann}, we get
\begin{align}
\Obs{\dlh} &= \tdif\Obs{B} \approx \tdif \left( 4\Ht\Obs{p_E} \right) \nonumber \\
&\approx \left( 4\frac{\dot{\Nb}}{\Nb}\Ht - 12\Ht^2 \right)\Obs{p_E} - \frac{\Nb^2}{A}\Obs{\phi} \nonumber \\
&= \left( \frac{\dot{\Nb}\Nb}{A\Ht} - 3\frac{\Nb^2}{A} \right) \Psi + \frac{\Nb^2}{A}\left( 2\Upsilon + \left( \frac{1}{2}+\frac{\kappa}{4}\frac{\Nb^2p}{\Ht^2} \right)\Psi \right) \label{SFGdlh}\ .
\end{align}

Since the vector sector is the same as in the longitudinal gauge, we also get
\be \Obs{\dl^a_\perp} = \left( 4\dot{\Ht} - 16\Ht^2+4\digamma\Ht \right)\nu^a \ . \ee

This completes the derivation of the Hamilton's equations of motions for all observables in the case of the spatially flat gauge.

\subsubsection{Second order equation for the Mukhanov--Sasaki variable}
We will show now that our formalism yields directly the evolution equation for the Mukhanov-Sasaki variable. We first notice that the Mukhanov-Sasaki variable $v= \Obs{\dphi}$ corresponds directly to the linearized observable of the scalar field perturbation $\dphi$ and we can therefore calculate its second time derivative directly via
\begin{align}
\ddot v &= \tdif \left( \tdif \Obs{\dphi} \right) \approx \tdif \left( \lphi\frac{\Nb\piphibar}{A^{\frac{3}{2}}} \Obs{\phi} + \lphi\frac{\Nb}{A^{\frac{3}{2}}}\piv\right)  \nonumber\\
&\approx \frac{\Nb^2}{A}\Delta v - \frac{\Nb^2}{2}V''(\bar\varphi) v + \left( \frac{\dot\Nb}{\Nb}-3\Ht \right)\dot v - \Nb^2V'(\bar\varphi)\Obs{\phi} + \lphi \frac{\Nb\piphibar}{A^{\frac{3}{2}}}\dotObs{\phi} \nonumber\\
&\quad\, + \lphi\frac{\Nb\piphibar}{A^{\frac{3}{2}}}\Delta \Obs{B} \ , \label{2ndMS}
\end{align}
using the previously derived Hamilton's equations for the linearized observables as well as
\begin{align}
\piv = \frac{A^{\frac{3}{2}}}{\Nb\lphi}\dot v - \piphibar \Obs{\phi} \ , \qquad  
 \dot{\pi}_v = \dotObs{\dpiphi} \ ,  \label{eq:piv}
\end{align}
~(\ref{eq:EOMmatter}) for the expression of $\dot{\bar\pi}_\varphi$ and \eqref{eq:Friedmann} for the expression of $\dot\Ht$.

Since we want to have a closed second order differential equation for $v$ only, we need to reformulate $\Obs{\phi}$, $\dotObs{\phi}$ and $\Obs{B}$ in terms of $v$ and $\dot v$. This is done in appendix~\ref{deriv:MS} in some detail and we only state the final result here:
\begin{align}
    {\ddot v} \approx \frac{\Nb^2}{A}\Delta v + \left( \frac{\dot{\Nb}}{\Nb}-3\Ht \right) {\dot v} + \left( \frac{\kappa^2}{8} \frac{\dot{\bar{\varphi}}^4}{\lphi^2\Ht^2} - \frac{3\kappa}{2} \frac{\dot{\bar{\varphi}}^2}{\lphi} - \frac{\kappa}{2} \frac{\Nb^2\dot{\bar{\varphi}}}{\lphi\Ht} V'(\bar\varphi) - \frac{\Nb^2}{2}V''(\bar\varphi) \right) v~. \label{eq:MS}
\end{align}

By this, we see that -- besides the second order equation of motion for the Bardeen potential we found in the course of the longitudinal gauge -- we can also derive the Mukhanov-Sasaki equation in the framework of Dirac observables in the spatially flat gauge.

We end this section by first mentioning that~(\ref{eq:MS}) agrees with the Mukhanov-Sasaki equation Langlois derived in~\cite{Langlois} (cf.  (56) therein)\footnote{To compare (56) of~\cite{Langlois} with~(\ref{eq:MS}) above, one has to insert the Friedmann equations~(\ref{eq:Friedmann}) and $\ddot{\bar{\varphi}} = \frac{\dot\Nb \dot{\bar{\varphi}}}{\Nb}-\frac{\Nb^2}{2}V'(\bar{\varphi})-3\Ht\dot{\bar{\varphi}} $. Note that Langlois uses proper time, i.e. $\Nb=1$, and a different definition for the potential, $V = 2V_{\rm Langlois}$, as well as $\Delta = A\Delta_{\rm Langlois}$.}. Secondly, we note that $v$ is the remaining scalar degree of freedom in the spatially flat gauge, with the other two being covered by the perturbed metric's tensorial part, $\Obs{h^{TT}_{ab}}$, as in the longitudinal gauge.

\section{Conclusions}
\label{Sec:SummConcl}
The main goal of this manuscript was to understand the dynamics of gauge invariant quantities in canonical cosmological perturbation theory. The evolution equations for such quantities are well known in the conventional approaches to linearized cosmological perturbations, but were never obtained in the canonical approach for different choices of gauge invariant variables. The fundamental reason for this can be traced  to the lack of understanding of an appropriate phase space construction which allows an explicit equivalence with standard and covariant perturbation techniques. Traditionally, in canonical perturbation theory one works with a (reduced) ADM phase space which treats lapse and shift as Lagrange multipliers, thus freezing their essential properties as phase space variables. This serves as the main roadblock to establish equivalence with the standard approach based on perturbations of Einstein field equations and the covariant approach, and masks various novel properties of the canonical theory.   Recent developments in  \cite{Giesel:2017roz, Giesel:2018opa}, provided a platform to fill these gaps and answer above questions. In these works, an extended ADM phase space formulation of canonical perturbation theory using relational formalism was made available. In the extended phase space, lapse and shift are restored their phase space character, owing to which bridges between canonical approach and the conventional perturbation theory can be established. Our analysis, using the results in \cite{Giesel:2017roz, Giesel:2018opa}, derived for the first time the gauge invariant dynamical equations for gauge invariant quantities or the Dirac obervables in the canonical cosmological perturbation theory. An important gap between the canonical and standard approaches to perturbation theory, thus gets filled.

We focused on two of the most popular gauge choices in cosmological perturbation theory: the longitudinal gauge and the spatially flat gauge. We derived the Hamilton's gauge invariant equations of motion for the Bardeen potential and the Mukhanov-Sasaki variable and showed in section \ref{Sec:Applic} that they agree with the second order differential equation obtained in the Lagrangian framework. To obtain these equations, it was essential to generalize a proof presented in \cite{Pons3,Pons4} that relates the gauge invariant equations of motions for the Dirac observables to the gauge variant ones.   The reason for a generalization of this result is tied to the fact that class of gauge fixing conditions relevant in linear cosmological perturbation theory do not fall in the  specific class of gauge fixing conditions considered in \cite{Pons3,Pons4}. Using observable map, a generalized lemma was established in our work which allows more general gauge fixing conditions, and in particular those relevant for cosmological perturbation theory to be considered.  With this generalization at hand we were able to compute the corresponding equations of motions in a straight forward way. Our method not only demonstrates that canonical perturbation theory does indeed yield the same equations as obtained in standard approach using Lagrangian methods, but that such an exercise is straightforward and efficient in contrast to standard approaches. It is to be emphasized that for the application of lemma \ref{MainLemma} it was pivotal that the observable map was defined on the extended ADM phase space and lapse and shift are not treated as Lagrange multipliers. 

Furthermore,  lemma \ref{MainLemma} provides an efficient technique to compute the equations of motion for Dirac observables without knowing the algebra of the observables or their corresponding physical Hamiltonian explicitly for the class of gauge fixing conditions under consideration in this work. Given that we have shown agreement between our formalism and the conventional Lagrangian approach also at the dynamical level, a next natural step will be to take this work as a basis and go beyond the classical theory and discuss the corresponding quantum dynamics for the observables. In this case, the knowledge of the algebra of the observables as well as their physical Hamiltonian is crucial. If we consider the results obtained here in the context of a reduced phase space quantization, then the final physical degrees of freedom consist of one scalar degree (the Bardeen potential or the Mukhanov-Sasaki variable) and two tensorial degrees of freedom. As a consequence, the physical Hamiltonian will only depend on these degrees of freedom and yield directly the corresponding equations of motions. In this work, we have followed closely the conventional approach where one considers the dynamics of a set of observables that are not independent of each other, and then uses their relationships to obtain the final equations of motion for the physical degrees of freedom. Since we derive the equations of motion by means of lemma \ref{MainLemma}, we will analyze in a future work whether it allows us to rederive the corresponding physical Hamiltonian. It will be interesting to explore its connection with the properties of the observable algebra in this context. 

\section*{Acknowledgements}
P.S. is grateful to the Institute for Quantum Gravity at Friedrich-Alexander-Universit\"{a}t Erlangen-N\"{u}rnberg for its generous support and a warm hospitality at various stages of this work.
D.W. thanks the German Academic Scholarship Foundation for financial support.

\appendix
\section{Proof of lemma \ref{MainLemma} in the general case}
\label{Sec:GenProof}
In this section, we generalize the proof of lemma \ref{MainLemma} from the specialized case in the main text, where the phase space function $f$ was not allowed to depend on lapse and shift, to phase space functions that can depend on all phase space degrees of freedom. This will involve to use the more complicated form of the observable map suitable for lapse and shift degrees of freedom as well as generalizing lemma \ref{Lemma2} to this case. If we rewrite $\frac{\d {\cal O}_{f,T}}{\d t}$ in terms of the Hamiltonian vector fields $X_{\tilde{\mathfrak{C}}_I}\cdot f\coloneqq \{f,\tilde{\mathfrak{C}}_I\}$, we obtain
\begin{eqnarray}
\frac{\d \mathcal{O}_{f,T}}{\d t}
&\approx&
\sum\limits_{n=0}^{\infty} \frac{1}{n!} \int\mathrm{d}^3y_1...\int\mathrm{d}^3y_n\int\mathrm{d}^3y_{n+1} \nonumber \\
&& \mathfrak{G}^{I_1}(y_1)...\mathfrak{G}^{I_{n}}(y_{n})\frac{\partial\mathfrak{G}^{I_{n+1}}}{\partial t}(y_{n+1})  X_{\tilde{\mathfrak C}_{I_n}}\cdots X_{\tilde{\mathfrak C}_{I_1}} X_{\tilde{\mathfrak C}_{I_{n+1}}}\cdot f .
\end{eqnarray}
Similarly to the specalized case in the main text, in order to be able to proof lemma \ref{MainLemma} here, we need to show that we can pull the term $\frac{\partial\mathfrak{G}^{I_{n+1}}}{\partial t}$ inside the iterated Poisson bracket, that means writing it next to the most inner Hamiltonian vector field. For this purpose, we adjust lemma \ref{Lemma2} from the main text to our more general situation here. This reads
\begin{lemma}
\label{Lemma3}
For a gauge fixing condition of the form in (\ref{eq:GenGF}) that is allowed to depend on all phase variables except lapse function and shift vector degrees of freedom and that can have an explicit dependence on coordinates and for which the associated matrix ${\cal  A}^\mu_\nu\coloneqq \{T^\mu,C_\nu\}$ has the property that  $\frac{\partial{\cal A}^\mu_\nu}{\partial t}\approx 0$, we have for all $n\in\mathbf{N}$:
\begin{align}
\frac{\partial {\mathfrak G}^{I_{n+1}}}{\partial t} & (y_{n+1}) X_{\tilde{\mathfrak C}_{I_1}(y_1)} \cdots X_{\tilde{\mathfrak C}_{I_n}(y_n)}X_{\tilde{\mathfrak C}_{I_{n+1}}(y_{n+1})}\cdot f 
 \nonumber \\
 & \approx X_{\tilde{\mathfrak C}_{I_n}(y_n)}\cdots X_{\tilde{\mathfrak C}_{
  I_1}(y_1)}X_{\frac{\partial {\mathfrak G}^{I_{n+1}}}{\partial t}\tilde{\mathfrak C}_{I_{n+1}}(y_{n+1})}\cdot f.	
\end{align}	
\end{lemma}
Using the Leibniz rule, we get
\begin{align}
\frac{\partial {\mathfrak G}^{I_{n+1}}}{\partial t} & (y_{n+1})
 X_{\tilde{\mathfrak C}_{I_n}(y_n)}\cdots X_{\tilde{\mathfrak C}_{I_1}(y_1)}X_{\tilde{\mathfrak C}_{I_{n+1}}(y_{n+1})}\cdot f \nonumber \\
 &=
 X_{\tilde{\mathfrak C}_{I_n}(y_n)}\cdots X_{\tilde{\mathfrak C}_{I_1}(y_1)}X_{\frac{\partial {\mathfrak G}^{I_{n+1}}}{\partial t}\tilde{\mathfrak C}_{I_{n+1}}(y_{n+1})}\cdot f	\nonumber \\
& \quad\ - \left(X_{\tilde{\mathfrak C}_{I_n}(y_n)}\cdots X_{\tilde{\mathfrak C}_{I_1}(y_1)}\cdot\frac{\partial {\mathfrak G}^{I_{n+1}}}{\partial t}\right)
\left(X_{\tilde{\mathfrak C}_{I_{n+1}}}\cdot f\right)  \nonumber \\
&\quad\ - X_{\tilde{\mathfrak C}_{I_n}(y_n)}\cdots X_{\tilde{\mathfrak C}_{I_1}(y_1)}\tilde{\mathfrak C}_{I_{n+1}}(y_{n+1})X_{\frac{\partial {\mathfrak G}^{I_{n+1}}}{\partial t}(y_{n+1})}\cdot f .
 	\end{align}
Now, we have to show that the last two terms on the right hand side of the last equation weakly vanish. This will be automatically given if the following two results hold:
\begin{eqnarray}
\label{eq:ConIundIIGen}
(i) && 	X_{\tilde{\mathfrak C}_{I_k}(y_k)}\cdots X_{\tilde{\mathfrak C}_{I_1}(y_1)}\cdot \tilde{\mathfrak C}_{I_{n+1}}(y_{n+1})\approx 0,\quad \forall k\in\mathbf{N}_0\quad {\rm and}\nonumber \\
(ii) && X_{\tilde{\mathfrak C}_{I_k}(y_k)}\cdots X_{\tilde{\mathfrak C}_{I_1}(y_1)}\cdot\frac{\partial {\mathfrak G}^{I_{n+1}}}{\partial t}\approx 0,\quad \forall k\in\mathbf{N}.
\end{eqnarray}
Let us focus on $(i)$ first. For $k=0$, $(i)$ is just given by $\tilde{\mathfrak C}_{I_n}(y_n)\approx 0$ and hence obviously weakly vanishes. For $k>1$, we can use that $\{\tilde{\mathfrak C}_{I}(x),\tilde{\mathfrak C}_{J}(y)\}\approx 0$ as well as the first class property of the constraints $\tilde{\mathfrak C}_{I}$'s from which trivially follows that $(i)$ is satisfied for all $k\geq 1\in\mathbbm{N}$. Now, for the second condition, as before, we need to consider also the terms that weakly vanish at each order to analyze carefully whether they contribute to the final result. In this context, we have to discuss a subtlety for the generalized observable formula. In our companion paper \cite{Giesel:2017roz}, we closely follow \cite{Pons3,Pons4} and replace the generator of gauge transformations of the extended ADM phase space by a weakly equivalent one that has a much simpler form, see  (2.52) in section 2.b in \cite{Giesel:2017roz}. In our notation here, we can incorporate this as follows. The non-simplified gauge generator derived in \cite{Pons3,Pons4} can be expressed in terms of generators $\tilde{\mathfrak C}'_I$ whose relation to $\tilde{\mathfrak C}_I$ used in our observable map is given by
\begin{eqnarray}
\label{eq:CundCPrime}
 \tilde{\mathfrak C}_I(x) 
 &=&\tilde{\mathfrak C}'_I(x) +\int \d^3y_{n+2} \int \d^3y_{n+3}{\cal F}^{\rho\lambda}_I(y_{n+2},y_{n+3},x)C_\rho(y_{n+2})\Pi_\lambda(y_{n+3})  \nonumber\\
& =&\tilde{\mathfrak C}'_I(x) +\int \d^3y_{n+4} \int \d^3y_{n+5}\widetilde{\cal F}^{\rho\lambda}_I(y_{n+4},y_{n+5},x)\tilde{\tilde{C}}_\rho(y_{n+4})\tilde{\Pi}_\lambda(y_{n+5}) ,
\end{eqnarray}
with the appropriate definitions of ${\cal F}^{\rho\lambda}_I$ and $\widetilde{\cal F}^{\rho\lambda}_I$. Their explicit form can be easily determined from the formulas in section 2.b in \cite{Giesel:2017roz} but is not important for our further discussion. The only information we need is how this additional term depends on the secondary and primary constraints. Given the relation in (\ref{eq:CundCPrime}), we obtain
\begin{eqnarray}
\{{\mathfrak G}^I(x),\tilde{\mathfrak C}_J(y)\}
&=&
\{{\mathfrak G}^I(x),\tilde{\mathfrak C}'_J(y)\nonumber\\
&& + \int \d^3y_{n+4} \int \d^3y_{n+5}\widetilde{\cal F}^{\rho\lambda}_J(y_{n+4},y_{n+5},y)\tilde{\tilde{C}}_\rho(y_{n+4})\tilde{\Pi}_\lambda(y_{n+5}) \}\nonumber \\
&=&
\{{\mathfrak G}^I(x),\tilde{\mathfrak C}'_J(y)\} \nonumber \\
&&+
\int \d^3y_{n+4} \int \d^3y_{n+5}\{{\mathfrak G}^I(x),\widetilde{\cal F}^{\rho\lambda}_J(y_{n+4},y_{n+5},y)\tilde{\tilde{C}}_\rho(y_{n+4})\tilde{\Pi}_\lambda(y_{n+5}) \}\nonumber \\
&=&
-\delta^I_J\delta^{(3)}(x,y)+ \int \d^3y_{n+4}\{{\mathfrak G}^I(x),{\mathfrak B}^L_J(y_{n+4},y)\}{\mathfrak C}_L(y_{n+4})\nonumber \\
&&+
\int \d^3y_{n+4} \int \d^3y_{n+5}\{{\mathfrak G}^I(x),\widetilde{\cal F}^{\rho\lambda}_J(y_{n+4},y_{n+5},y) \}\tilde{\tilde{C}}_\rho(y_{n+4})\tilde{\Pi}_\lambda(y_{n+5})\nonumber \\
&&+
\int \d^3y_{n+4} \int \d^3y_{n+5}\{{\mathfrak G}^I(x),\tilde{\tilde{C}}_\rho(y_{n+4}) \}\widetilde{\cal F}^{\rho\lambda}_J(y_{n+4},y_{n+5},y)\tilde{\Pi}_\lambda(y_{n+5})\nonumber \\
&&+
\int \d^3y_{n+4} \int \d^3y_{n+5}\{{\mathfrak G}^I(x),\tilde{\Pi}_\lambda(y_{n+5}) \}\widetilde{\cal F}^{\rho\lambda}_J(y_{n+4},y_{n+5},y)\tilde{\tilde{C}}_\rho(y_{n+4}).\nonumber \\
\end{eqnarray}
We realize that all terms but the first one are at least linear in the primary and secondary constraints respectively 
which will be crucial for the next step where we consider the temporal partial derivative of $\mathfrak{G}^I$. This leads to
\begin{eqnarray}
&& \hspace{-1.5cm}\{\frac{\partial {\mathfrak G}^I}{\partial t}(x),\tilde{\mathfrak C}_J(y)\}  \nonumber\\
&=&
\frac{\partial}{\partial t}\left(\{{\mathfrak G}^I(x),\tilde{\mathfrak C}_J(y)\}\right)
-\{{\mathfrak G}^I(x),\frac{\partial\tilde{\mathfrak C}_J}{\partial t}(y)\}\nonumber \\
&=&
\int d^3y_{n+4}\int \d^3y_{n+5}\{\frac{\partial}{\partial t}{\mathfrak G}^I(x),{\mathfrak B}^L_J(y_{n+4},y)\}{\mathfrak A}_L^N(y_{n+4},y_{n+5})\tilde{\mathfrak C}_N(y_{n+5})
\nonumber \\
&&-
\int \d^3y_{n+4}\{{\mathfrak G}^I(x),{\mathfrak C}'_L(y_{n+4})\}\frac{\partial}{\partial t}{\mathfrak B}^L_J(y_{n+4},y)\nonumber \\
&&-
\int \d^3y_{n+4}\int \d^3y_{n+5}\int \d^3y_{n+6} \frac{\partial }{\partial t}{\mathfrak B}^L_J(y_{n+4},y)
 \nonumber\\
&& \hphantom{-\int} \{ {\mathfrak G}^I(x), \tilde{\tilde{{\cal F}}}^{\rho\lambda}_L (y_{n+5},y_{n+6},y_{n+4})
\tilde{\tilde{C}}_\rho (y_{n+5})\tilde{\Pi}_\lambda(y_{n+6}) \}
\nonumber \\
&&+
\frac{\partial }{\partial t}\int \d^3y_{n+4} \int \d^3y_{n+5}\{{\mathfrak G}^I(x),\widetilde{\cal F}^{\rho\lambda}_J(y_{n+4},y_{n+5},y) \}\tilde{\tilde{C}}_\rho(y_{n+4})\tilde{\Pi}_\lambda(y_{n+5})\nonumber \\
&&+
\frac{\partial }{\partial t}\int \d^3y_{n+4} \int \d^3y_{n+5}\{{\mathfrak G}^I(x),\tilde{\tilde{C}}_\rho(y_{n+4}) \}\widetilde{\cal F}^{\rho\lambda}_J(y_{n+4},y_{n+5},y)\tilde{\Pi}_\lambda(y_{n+5})\nonumber \\
&&+
\frac{\partial }{\partial t}\int \d^3y_{n+4} \int \d^3y_{n+5}\{{\mathfrak G}^I(x),\tilde{\Pi}_\lambda(y_{n+5}) \}\widetilde{\cal F}^{\rho\lambda}_J(y_{n+4},y_{n+5},y)\tilde{\tilde{C}}_\rho(y_{n+4})
\end{eqnarray}
and further
\begin{eqnarray}
\label{eq:InnerPBGen}
&& \hspace{-1.5cm}\{\frac{\partial {\mathfrak G}^I}{\partial t}(x),\tilde{\mathfrak C}_J(y)\}  \nonumber\\
&=&
\int \d^3y_{n+4}\int \d^3y_{n+5}\{\frac{\partial}{\partial t}{\mathfrak G}^I(x),{\mathfrak B}_J^L(y_{n+4},y)\}{\mathfrak A}_L^N(y_{n+4},y_{n+5})\tilde{\mathfrak C}_N(y_{n+5})
\nonumber \\
&&+
\int \d^3y_{n+4}\mathfrak{A}^I_L(x,y_{n+4})\frac{\partial }{\partial t}{\mathfrak B}^L_J(y_{n+4},y)
\nonumber \\
&&-
\int \d^3y_{n+4}\int \d^3y_{n+5}\int \d^3y_{n+6} \frac{\partial }{\partial t}{\mathfrak B}^L_J(y_{n+4},y)
 \nonumber\\
&& \hphantom{-\int} \{ {\mathfrak G}^I(x), \tilde{\tilde{{\cal F}}}^{\rho\lambda}_L (y_{n+5},y_{n+6},y_{n+4})
\tilde{\tilde{C}}_\rho (y_{n+5})\tilde{\Pi}_\lambda(y_{n+6}) \}
\nonumber \\
&&+
\frac{\partial }{\partial t}\int \d^3y_{n+4} \int \d^3y_{n+5}\{{\mathfrak G}^I(x),\widetilde{\cal F}^{\rho\lambda}_J(y_{n+4},y_{n+5},y) \}\tilde{\tilde{C}}_\rho(y_{n+4})\tilde{\Pi}_\lambda(y_{n+5})\nonumber \\
&&+
\frac{\partial }{\partial t}\int \d^3y_{n+4} \int \d^3y_{n+5}\{{\mathfrak G}^I(x),\tilde{\tilde{C}}_\rho(y_{n+4}) \}\widetilde{\cal F}^{\rho\lambda}_J(y_{n+4},y_{n+5},y)\tilde{\Pi}_\lambda(y_{n+5})\nonumber \\
&&+
\frac{\partial }{\partial t}\int \d^3y_{n+4} \int \d^3y_{n+5}\{{\mathfrak G}^I(x),\tilde{\Pi}_\lambda(y_{n+5}) \}\widetilde{\cal F}^{\rho\lambda}_J(y_{n+4},y_{n+5},y)\tilde{\tilde{C}}_\rho(y_{n+4}),
\end{eqnarray}
where we introduced $ \tilde{\tilde{{\cal F}}}^{\rho\lambda}_L$ with the appropriate definition in the fourth line so that the equality is satisfied.
If we compare this to the result in (\ref{eq:InnerPB}) and using $G^\mu=-T^\mu$, we realize that the first two terms here are of the same kind as the two terms in (\ref{eq:InnerPB}). The additional terms in the result in (\ref{eq:InnerPBGen}) are caused by the fact that we have to involve also all weakly vanishing terms that come from the replacement of the gauge generator by a weakly equivalent one up to terms quadratic in the constraints. Our next step consists in writing condition $(ii)$ in (\ref{eq:ConIundIIGen}) in terms of the Poisson bracket computed in (\ref{eq:InnerPBGen}). We end up with
\begin{equation}
  X_{\tilde{\mathfrak C}_{I_k}(y_k)}\cdots X_{\tilde{\mathfrak C}_{I_1}(y_1)}\cdot\frac{\partial {\mathfrak G}^{I_{n+1}}}{\partial t} (y_{n+1})
  =X_{\tilde{\mathfrak C}_{I_k}(y_k)}\cdots X_{\tilde{\mathfrak C}_{I_{2}}(y_{2})}\cdot\{\frac{\partial {\mathfrak G}^{I_{n+1}}}{\partial t}(y_{n+1}),\tilde{\mathfrak C}_{I_1}(y_1)\} .
\end{equation}
As in the specialized case in the main text, we have to show that all terms on the right hand side of (\ref{eq:InnerPBGen}) weakly vanish for all values of $k\in \mathbf{N}$. The first and the third term in (\ref{eq:InnerPBGen}) are obviously linear in the constraints. If we apply the partial temporal derivative onto the last three terms, each resulting term can also again be written as expressions that involve the constraints at least linearly. Hence, for the case of $k=1$ all these terms weakly vanish. For $k>1$, we use the fact that each of this terms involves the constraints at least linearly together with $(i)$, and we can easily conclude that also for all $k>1$ the contributions of these terms vanish weakly.
The only remaining term is the second one in (\ref{eq:InnerPBGen}). Here, we consider our assumption in lemma \ref{MainLemma}, namely that $\partial_t{\cal A}^\mu_\nu\approx 0$. Considering the explicit form of ${\mathfrak A}^J_K$ in (\ref{eq:AMatrix}), we want to show that using $\partial_t{\cal A}^\mu_\nu\approx 0$ we can conclude $\partial_t{\mathfrak A}^I_J\approx 0$. Thus, we need to prove that from $\partial_t{\cal A}^\mu_\nu\approx 0$ follows $\partial_t \{\dot{T}^\mu,C_\nu\}\approx 0$. If we use the Jacobi identity and ${\cal A}^\mu_\nu=\{T^\mu,C_\nu\}$, we obtain
\begin{equation}
\label{eq:PartialPB}
\frac{\partial }{\partial t}\{\dot{T}^\mu,C_\nu\}
=
\frac{\partial}{\partial t}\frac{\partial}{\partial t} {\cal A}^{\mu}_{\nu}+N^\lambda C^\rho_{\lambda\nu}\frac{\partial}{\partial t}{\cal A}^{\mu}_\rho-N^\lambda\{\frac{\partial}{\partial t}{\cal A}^{\mu}_\nu,C_\lambda\},   
\end{equation}
where $C^\rho_{\lambda\nu}$ denote the structure constants of the secondary constraints $C_\mu$. By our assumption, $\partial_t{\cal A}^\mu_\nu\approx 0$ vanishes at least weakly. If it vanishes strongly, then we also have $\partial_t{\mathfrak A}^I_J= 0$. In case it vanishes weakly, it can be written as a linear combination of the primary and secondary constraints, that is $\partial_t{\cal A}^\mu_\nu=\alpha^{\mu\gamma}_\nu C_\gamma+\beta^{\mu\sigma}_\nu \Pi_\sigma$ for appropriate choices of the coefficients $\alpha,\beta$. Reinserting this into the last equation and taking into account that the constraints are first class, we get $\frac{\partial }{\partial t}\{\dot{T}^\mu,C_\nu\}\approx 0$. As a consequence, we obtain $\partial_t{\mathfrak A}^I_J\approx 0$. Because $\partial_t{\mathfrak B}^I_J$ is the inverse of ${\mathfrak A}^I_J$, this carries over to $\partial_t{\mathfrak B}^I_J\approx 0$ by means of the Leibniz rule. Therefore, for $k=1$, this term at least weakly vanishes by assumption. For larger $k>1$, we can again use $(i)$ because either $\partial_t{\mathfrak B}^I_J\approx 0$ vanishes already strongly, then we do not need to take this into account for any value of $k$, or it only weakly vanishes, but then it can be written as a linear combination of the $\tilde{\mathfrak C}_I$ and, with $(i)$, its contribution for all $k\in\mathbf{N}$ weakly vanishes. Thus, we have proven lemma \ref{Lemma3} for the general case involving lapse and shift degrees of freedom. Now, using lemma \ref{Lemma3}, the time derivative of the observable can be rewritten as
\begin{eqnarray}
\frac{\d \mathcal{O}_{f,T}}{\d t}
&\approx&
\sum\limits_{n=0}^{\infty} \frac{1}{n!} \int\mathrm{d}^3y_1...\int\mathrm{d}^3y_n
 \mathfrak{G}^{I_1}(y_1)...\mathfrak{G}^{I_{n}}(y_{n}) \nonumber \\
&& X_{\tilde{\mathfrak C}_{I_n}}\cdots X_{\tilde{\mathfrak C}_{I_1}} X_{\int\mathrm{d}^3y_{n+1}\frac{\partial\mathfrak{G}^{I_{n+1}}}{\partial t}(y_{n+1})\tilde{\mathfrak C}_{I_{n+1}}}\cdot f.
\end{eqnarray}
As our final step, we will show that the most inner Hamiltonian vector field can be rewritten as the Hamiltonian vector field associated with $H_{\rm can}$. We have
\begin{align}
\int\mathrm{d}^3y_{n+1} & \frac{\partial\mathfrak{G}^{I_{n+1}}}{\partial t}(y_{n+1})\tilde{\mathfrak C}_{I_{n+1}} \nonumber\\
&= \int \d^3y_{n+1}\frac{\partial G^{\mu_{n+1}}}{\partial t}\tilde{\tilde{C}}_{\mu_{n+1}}(y_{n+1})
+\int \d^3y_{n+1}\frac{\partial}{\partial t} \frac{\d G^{\mu_{n+1}}}{\d t}\tilde{\Pi}_{\mu_{n+1}}(y_{n+1}) .
\end{align}
Considering (\ref{eq:RelationConst}), we can rewrite the constraints $\tilde{\tilde{C}}_\mu$ and $\tilde{\Pi}_\mu$ in terms of the original constraints $C_\mu$ and $\Pi_\mu$ yielding
\begin{align}
\int \d^3& y_{n+1} \frac{\partial \mathfrak{G}^{\mu_{n+1}}}{\partial t}\tilde{\tilde{\mathfrak{C}}}_{\mu_{n+1}}(y_{n+1}) \nonumber\\
& \approx \int \d^3y_{n+1}\int \d^3y_{n+2}\frac{\partial G^{\mu_{n+1}}(y_{n+1})}{\partial t}{\cal B}^{\mu_{n+2}}_{\mu_{n+1}}(y_{n+2},y_{n+1})C_{\mu_{n+2}}(y_{n+2})\nonumber \\
&\quad\ -
\int \d^3y_{n+1}\cdots \int \d^3y_{n+4}\frac{\partial G^{\mu_{n+1}}(y_{n+1})}{\partial t}{\cal B}^{\mu_{n+2}}_{\mu_{n+1}}(y_{n+2},y_{n+1}){\cal B}^{\mu_{n+4}}_{\mu_{n+3}}(y_{n+4},y_{n+3})\nonumber \\
&\hphantom{-\int}\quad
\{\dot{T}^{\mu_{n+3}}(y_{n+3}),C_{\mu_{n+2}}(y_{n+2})\}\Pi_{\mu_{n+4}}(y_{n+4})\nonumber \\
&\quad\ 
+\int \d^3y_{n+1}\int \d^3y_{n+2}\frac{\partial}{\partial t} \frac{\d G^{\mu_{n+1}}(y_{n+1})}{\d t}{\cal B}^{\mu_{n+2}}_{\mu_{n+1}}(y_{n+2},y_{n+1})\Pi_{\mu_{n+2}}(y_{n+2}) .
\end{align}
We consider the expression for $N^\mu$ in (\ref{eq:GmuStab}) and for the Langrage multiplier $\lambda^\mu$ in (\ref{eq:FixLambda}) and realize that the first integral on the right hand side of the last equation involve $N^\mu$ whereas the second and the third integral combine exactly to $\lambda^\mu$. Thus, we obtain 
\begin{align}
\int\mathrm{d}^3&y_{n+1}\frac{\partial\mathfrak{G}^{I_{n+1}}}{\partial t}(y_{n+1})\tilde{\mathfrak C}_{I_{n+1}} \nonumber\\
&\approx
\int \d^3y_{n+1}\left(N^{\mu_{n+1}}C_{\mu_{n+1}}+\lambda^{\mu_{n+1}}\Pi_{\mu_{n+1}}\right)(y_{n+1})\nonumber \\
&=
H_{\rm can} .
\end{align}
We reinsert this into the total time derivative of the observable to finally get
\begin{eqnarray}
\frac{\d \mathcal{O}_{f,T}}{\d t}
&\approx&
\sum\limits_{n=0}^{\infty} \frac{1}{n!} \int\mathrm{d}^3y_1...\int\mathrm{d}^3y_n
\left( \mathfrak{G}^{I_1}(y_1)...\mathfrak{G}^{I_{n}}(y_{n})
  X_{\tilde{\mathfrak C}_{I_n}}\cdots X_{\tilde{\mathfrak C}_{I_1}} X_{H_{\rm can}}\cdot f\right)\nonumber \\
&=&
\sum\limits_{n=0}^{\infty} \frac{1}{n!} \int\mathrm{d}^3y_1...\int\mathrm{d}^3y_n
\left( \mathfrak{G}^{I_1}(y_1)...\mathfrak{G}^{I_{n}}(y_{n}) X_{\tilde{\mathfrak C}_{I_n}}\cdots X_{\tilde{\mathfrak C}_{I_1}}\cdot \{f,H_{\rm can}\}\right)\nonumber \\
&=&
\mathcal{O}_{\{f,H_{\rm can}\},T} .
\end{eqnarray}
This finishes our proof of lemma \ref{MainLemma} for the general case and as our presentation shows the proof in the general case can be formulated along the lines of the specialized case.

\section{Towards the final Bardeen equation}\label{deriv:Bardeen}

The purpose of this section is to illustrate how to obtain the final second order equation of motion for the Bardeen potential,~(\ref{eq:Bardeen}), from the starting point~(\ref{eq:StartBardeen}):
\be
\ddot \Psi \approx \left(\frac{\dot{\bar N}}{\bar N} - 4 \Ht \right) \dot \Psi + \frac{\kappa}{2} \bar N^2 p \Psi - \frac{\kappa}{4} \bar N^2 \dT ~. \tag{\ref{eq:StartBardeen}}
\ee

We first recap~(C8) of~\cite{Giesel:2017roz}\footnote{For reasons of being concise, we now refer to equations of~\cite{Giesel:2017roz} by ``\paperref{\ldots}'' whenever we use one.} for the gauge invariant extension of~(\ref{eq:deltaT}):
\begin{equation}
 \dT = -3\lphi\frac{\piphibar^2}{A^3}\Psi + \lphi\frac{\piphibar}{A^3}\delta\pi_\varphi\gi - \frac{1}{2\lphi}V'(\bar\varphi)\delta\varphi\gi ~.    \label{eq:dT}
\end{equation}
So we need expressions for $\delta\pi_\varphi\gi$ and $\delta\varphi\gi$ in terms of $\Psi$, which we can obtain via the components of $\dT$. Starting with the second equation of~\appref{34}, we get
\begin{align}
2\Ht\left( \Upsilon - \frac{1}{2}\Psi \right)_{,a} & \approx \frac{\kappa}{4}\Nb^2 \delta{\tilde T}^0_a{}\gi \nonumber\\
& \hspace{-.35cm}\overset{\appref{31}}{=} \frac{\kappa}{4}\Nb^2\delta{\tilde T}^0_a - \kappa\lphi\frac{\Ht\piphibar^2}{A^2} \left( \Obs{E}+\Obs{p_E} \right)_{,a} \nonumber \\
&  \hspace{-.35cm}\overset{\appref{23}}{=} -\frac{\kappa}{4}\frac{\Nb\piphibar}{A^{\nicefrac{3}{2}}} \left(\Obs{\delta\varphi}\right)_{,a}~,
\end{align}
where we also used that $\Obs{E}=\Obs{p_E}=0$ in the longitudinal gauge. We can therefore identify the left with the right hand side without the differentiation respectively and obtain for $\delta\varphi\gi$ by inserting~\paperref{3.123}
\begin{eqnarray}
\delta\varphi\gi \approx -\frac{8}{\kappa}\frac{\Ht A^{\nicefrac{3}{2}}}{\Nb\piphibar}\left( \Upsilon - \frac{1}{2}\Psi \right)~. \label{eq:dphigi}
\end{eqnarray}

Next, we want to express $\delta\pi_\varphi\gi$ in terms of $\Psi$ only and begin with the first equation of~\appref{34}:
\begin{align}
3\Ht^2\left(\Psi-2\Upsilon\right) + \frac{\Nb^2}{A}\Delta\Psi & \approx \frac{\kappa}{4} \Nb^2\delta{\tilde T}^0_0{}\gi \nonumber\\
& \hspace{-.35cm}\overset{\appref{27}}{=} \frac{\kappa}{4}\Nb^2\delta{\tilde T}^0_0+ 3\kappa \lphi\frac{\Ht^2\piphibar^2}{A^2}\left( \Obs{E}+\Obs{p_E} \right) \nonumber\\
& \hspace{-.35cm}\overset{\appref{22}}{=}  \frac{\kappa}{4}\Nb^2\left( 3\lphi\frac{\piphibar^2}{A^3}\Obs{\psi} - \lphi\frac{\piphibar}{A^3}\Obs{\delta\pi_\varphi} - \frac{1}{2\lphi}V'(\bar\varphi)\Obs{\delta\varphi} \right)~.
\end{align}
With $\Obs{\psi}=\Psi$ and~\paperref{3.123} and~\paperref{3.124}, we obtain
\begin{align}
    \delta\pi_\varphi\gi \approx 3\piphibar\Psi - \frac{4}{\kappa\lphi}\frac{A^2}{\piphibar}\Delta\Psi + \frac{4}{\kappa\lphi}\frac{A^3}{\Nb^2\piphibar} \left( \frac{1}{\lphi}\frac{\Nb\Ht A^{\nicefrac{3}{2}}}{\piphibar} V'(\bar\varphi) + 6\Ht^2 \right)\left(\Upsilon-\frac{1}{2}\Psi\right)~. \label{eq:dpiphigi}
\end{align}
Now,~(\ref{eq:dphigi}) and~(\ref{eq:dpiphigi}) together with~\appref{33}, $\Upsilon = \frac{1}{2\Ht}\dot\Psi + \Psi$, can be combined and put into~(\ref{eq:dT}) to get
\begin{align}
    \dT \approx - \frac{4}{\kappa A}\Delta \Psi + \left( \frac{4}{\kappa\lphi}\frac{\Ht A^{\nicefrac{3}{2}}}{\Nb\piphibar} V'(\bar\varphi) + \frac{12}{\kappa}\frac{\Ht^2}{\Nb^2} \right)\left( \frac{1}{\Ht}\dot\Psi + \Psi \right) ~.
\end{align}
Inserting this into~(\ref{eq:StartBardeen}) results in~(\ref{eq:Bardeen}) after also using~(\ref{eq:EOMmatter}) for $\piphibar$.

\section{Towards the final Mukhanov-Sasaki equation}\label{deriv:MS}

Analogously to the section before, we now want to present the essential steps for obtaining our final second order equation of motion for the Mukhanov-Sasaki variable,~(\ref{eq:MS}), when starting with~(\ref{2ndMS}),
\begin{align}
\ddot v & \approx \frac{\Nb^2}{A}\Delta v - \frac{\Nb^2}{2}V''(\bar\varphi) v + \left( \frac{\dot\Nb}{\Nb}-3\Ht \right)\dot v - \Nb^2V'(\bar\varphi)\Obs{\phi} + \lphi \frac{\Nb\piphibar}{A^{\frac{3}{2}}}\dotObs{\phi} \nonumber\\
&\quad\, + \lphi\frac{\Nb\piphibar}{A^{\frac{3}{2}}}\Delta \Obs{B}~. \tag{\ref{2ndMS}}
\end{align}
Since we want to have a second order equation in $v$ alone, we need to reformulate $\Obs{\phi}$, $\dotObs{\phi}$ and $\Obs{B}$ in terms of $v$ and $\dot v$. To get an expression in terms of $v$ for $\Obs{\phi}$, we use the observable of $\delta{\hat C}$ -- the scalar part of the diffeomorphism constraint:
\begin{align}
    \Obs{\delta{\hat C}} & = -4 A {\tilde P} \left(\Obs{p_\psi} + \frac{2}{3}\Delta \Obs{p_E}\right) + \kappa\piphibar\Obs{\delta\varphi} \nonumber\\
    & \approx 2A{\tilde P}\Obs{\phi} + \kappa\piphibar v \approx 0 \ .
\end{align}
In doing so, we used the stability condition of $\psi \approx 0$,~(\ref{eq:stabilitySFG}), on the observable level:
\begin{align}
    \Obs{\phi} \approx -2\Obs{p_\psi} - \frac{4}{3}\Delta\Obs{p_E} \ .
\end{align}
Thus, we finally get our desired expressions
\begin{align}
    \Obs{\phi} &\approx \frac{\kappa}{4} \frac{\Nb\piphibar}{A^{\frac{3}{2}}\Ht} v \quad \text{and, derived thereof,} \label{eq:Ophi} \\
    \dotObs{\phi} &\approx \left( \frac{\kappa^2}{16} \frac{\Nb^3\piphibar p}{A^\frac{3}{2}\Ht^2} - \frac{\kappa}{8}\frac{\Nb^2}{\lphi\Ht}V'(\bar\varphi) - \frac{3\kappa}{8}\frac{\Nb\piphibar}{A^\frac{3}{2}}\right)v + \frac{\kappa}{4}\frac{\Nb\piphibar}{A^\frac{3}{2}\Ht} \,{\dot v} \ . \label{eq:dotOphiMS}
\end{align}
Finding an expression in terms of $v$ and $\dot v$ for $\Obs{B}$ turns out to be more elaborate. We start with $\Obs{B} = \frac{\Nb^2}{A\Ht}\Psi $, which implies 
\begin{align}
     \Delta \Obs{B}=\frac{\Nb^2}{A\Ht}\Delta\Psi~, \label{eq:DeltaOB}
\end{align}
and then use
\begin{align}
    \Delta\Psi = \frac{\kappa}{4}A\,\delta T^0_0{}^{\rm (gi)} + 3\frac{A\Ht}{\Nb^2}\left(\Ht\Psi + \dot\Psi\right) \label{eq:DeltaPsi}
\end{align}
to continue with finding a respective expression for $\Delta\Psi$. The origin of~(\ref{eq:DeltaPsi}) can be found in~\cite{Giesel:2017roz},~\appref{28}.

Now, for expressing the last bracket of~(\ref{eq:DeltaPsi}) in terms of $v$, we continue with~\appref{32}:
\begin{align}
    \left( \Ht\Psi + {\dot\Psi} \right)_{,a} &= \frac{\kappa}{4}\Nb^2 \delta T^0_a{}^{\rm (gi)} \nonumber\\
    & \hspace{-.35cm} \overset{\appref{31}}{=} \frac{\kappa}{4}\Nb^2 \left( \delta T^0_a - \lphi\frac{4\Ht\piphibar^2}{\Nb^2A^2}\left( \Obs{E}+\Obs{p_E} \right)_{,a} \right) \nonumber\\
    & \hspace{-.35cm} \overset{\appref{23}}{\approx} \frac{\kappa}{4}\Nb^2 \left( -\frac{\piphibar}{\Nb A^\frac{3}{2}}\left(\Obs{\delta\varphi}\right)_{,a} - \lphi\frac{4\Ht\piphibar^2}{\Nb^2A^2}\left(\Obs{p_E}\right)_{,a} \right) \ ,
\end{align}
where we used $\Obs{E}\approx0$. This leads to
\begin{align}
    \Ht\Psi + {\dot\Psi} \approx -\frac{\kappa}{4} \frac{\Nb\piphibar}{A^\frac{3}{2}}v - \kappa\lphi\frac{\Ht\piphibar^2}{A^2}\Obs{p_E} \ . \label{eq:dotPsiMS}
\end{align}

Finally, we use \appref{27} and \appref{24} to reformulate
\begin{align}
    \delta T^0_0{}^{\rm (gi)} &= \delta T^0_0 + 12\lphi\frac{\Ht^2\piphibar^2}{A^2\Nb^2}\left( \Obs{E}+\Obs{p_E} \right) \nonumber\\
    & \approx -\lphi\frac{\piphibar}{A^3}\Obs{\delta\pi_\varphi} - \frac{1}{2\lphi}V'(\bar\varphi) v + 12\lphi\frac{\Ht^2\piphibar^2}{A^2\Nb^2}\Obs{p_E} \ .
\end{align}
Plugging in the observable equivalent of~(\ref{eq:piv}), recap that $\Obs{\delta\pi_\varphi}=\piv$, we get
\begin{align}
    \Obs{\delta\pi_\varphi} = \frac{A^\frac{3}{2}}{\lphi\Nb} {\dot v} - \piphibar\Obs{\phi}
\end{align}
and combining this with~(\ref{eq:dotPsiMS}) and~(\ref{eq:DeltaPsi}), formula~(\ref{eq:DeltaOB}) becomes
\begin{align}
    \Delta\Obs{B} \approx -\frac{\kappa}{4}\frac{\Nb\piphibar}{A^\frac{3}{2}\Ht} {\dot v} - \left( \frac{\kappa}{8}\frac{\Nb^2}{\lphi\Ht}V'(\bar\varphi) + \frac{3\kappa}{4}\frac{\Nb\piphibar}{A^\frac{3}{2}} \right)v + \frac{\kappa}{4}\frac{\lphi\piphibar^2\Nb^2}{A^3\Ht}\Obs{\phi} \ ,
\end{align}
where we notice that the terms proportional to $\Obs{p_E}$ indeed cancel each other.

 We can therefore now combine this expression with the ones for $\Obs{\phi}$ and $\dotObs{\phi}$,~(\ref{eq:Ophi}) and~(\ref{eq:dotOphiMS}), in order to obtain the Mukhanov-Sasaki equation from our previous~(\ref{2ndMS}):
\begin{align}
    {\ddot v} - & \frac{\Nb^2}{A}\Delta v + \frac{\Nb^2}{2}V''(\bar\varphi)v - \left( \frac{\dot{\Nb}}{\Nb}-3\Ht \right) {\dot v}  \nonumber\\
    & \approx -\frac{\kappa}{2}\frac{\Nb^3\piphibar}{A^\frac{3}{2}\Ht}V'(\bar\varphi)v - \frac{9\kappa}{8}\lphi\frac{\Nb^2\piphibar^2}{A^3}v + \frac{\kappa^2}{16}\lphi\frac{\Nb^4\piphibar^2}{A^3\Ht^2}\left(p+\lphi\frac{\piphibar^2}{A^3}\right)v \nonumber\\
    & = -\frac{\kappa}{2} \frac{\Nb^2\dot{\bar{\varphi}}}{\lphi\Ht} V'(\bar\varphi)v - \frac{3\kappa}{2} \frac{\dot{\bar{\varphi}}^2}{\lphi} v +\frac{\kappa^2}{8} \frac{\dot{\bar{\varphi}}^4}{\lphi^2\Ht^2} v \ ,
\end{align}
where we used $\lphi\frac{\piphibar^2}{A^3} = p+\rho $ (cf.~(\ref{eq:prho})), the Friedmann equation $\rho = \frac{6\Ht^2}{\kappa\Nb^2}$ and~(\ref{eq:EOMmatter}) for the expression of $\piphibar$ for going from the second to the third line.

\section{Dirac observables in perturbation theory}
\label{sec:DiracObsPert}
 In this appendix, we briefly discuss how the condition that Dirac observables at least weakly commute with all constraints in the full non-linear theory can be carried over to perturbation theory. The discussion is in the setting of extended ADM phase space.
 
 For this purpose, let us consider an extended ADM phase space with elementary variables $(q_{ab}(x),P^{ab}(x),N^\mu(x),\Pi_\mu(x))$, first class constraints $C_\mu(x)\approx 0$ and $\Pi_\mu(x)\approx 0$, with $\mu=0,\ldots,3$, and a corresponding canonical Hamiltonian $H_{\rm can}=\int \d^3x \left(N^\mu C_\mu+\lambda^\mu \Pi_\mu\right)(x)$, where $\lambda^\mu$ are Lagrange multipliers for $\mu=0,\ldots,3$. As presented in detail in \cite{Giesel:2017roz,Giesel:2018opa} and following \cite{Pons3,Pons4}, the generator of diffeomorphisms on the extended ADM phase space is given by
\begin{equation*}
G_\xi=\int \d^3x\left(\dot{\xi}^\mu\tilde{\Pi}_\mu+\xi^\mu\tilde{\tilde{C}}_\mu\right)(x),    
\end{equation*}
where $\dot{\xi}^\mu, \xi^\mu$ are arbitrary descriptors and $\tilde{\Pi}_\mu, \tilde{\tilde{C}}_\mu$ are weakly equivalent forms of the first class constraints $\Pi_\mu, C_\mu$. On the extended ADM phase space, the condition for weak  Dirac observables ${\cal O}_f$ reads
\begin{equation*}
\{{\cal O}_f,G_\xi\}\approx 0.
\end{equation*}
Since this condition needs to be satisfied for any chosen descriptors, it is equivalent to,
\begin{equation*}
\{{\cal O}_f,\tilde{\tilde{C}}_\mu\}\approx \{{\cal O}_f,C_\mu\}\approx 0, \quad 
\{{\cal O}_f,\tilde{\Pi}_\mu\}\approx \{{\cal O}_f,\Pi_\mu\}\approx 0.  
\end{equation*}
We wish to address the question the way this condition for observables carries over to the case of perturbation theory and how this is related to the evolution of observables. To answer this question, we generalize the discussion in \cite{Giesel:2007wi,Giesel:2007wk}, where in appendix E of \cite{Giesel:2007wi} the notion of constants of motion in perturbation theory was analyzed for Hamiltonian systems. Similar to the presentation in \cite{Giesel:2007wi}, we consider a finite dimensional analog whose results can be straight forwardly generalized to the field theoretic context relevant for the ADM case. Hence, we consider the simplified model of an extended phase space with elementary variables $(q,p,N^\mu,\Pi_\mu)$, with $\mu=0,\ldots,3$, satisfying the canonical Poisson brackets $\{q,p\}=1$ and $\{N^\mu,\Pi_\nu\}=\delta^\mu_\nu$, where all remaining Poisson brackets vanish. The eight constraints are of the form $C_\mu(q,p)\approx 0$ and $\Pi_\mu\approx 0$ and the canonical Hamiltonian is given by $H_{\rm can}=N^\mu C_\mu+\lambda^\mu\Pi_\mu$. The corresponding equations of motion induced by $H_{\rm can}$ read
\begin{eqnarray}
\frac{\d q}{\d t}&=&\{q,H_{\rm can}\},\quad\hspace{1,6cm} \frac{\d q}{\d t}=\{q,H_{\rm can}\}\nonumber\\
\frac{\d N^\mu}{\d t} &=& \{N^\mu,H_{\rm can}\}=\lambda^\mu,\quad \frac{\d \Pi_\mu}{\d t} = \{\Pi_\mu,H_{\rm can}\}=-C_\mu\nonumber .\\
\end{eqnarray}
Let $(\overline{q},\overline{p},\overline{N}^\mu,\overline{\Pi}_\mu)$ be an exact solution of the dynamical system above. We consider perturbations around this background solution of the form
\begin{equation*}
    \delta q=q-\overline{q},\quad \delta p=p-\overline{p},\quad \delta N^\mu=N^\mu-\overline{N}^\mu, \quad \delta\Pi_\mu=\Pi_\mu-\overline{\Pi}_\mu.
\end{equation*}
These have the following non-vanishing Poisson brackets:  $\{\delta q,\delta p\}=1$ and $\{\delta N^\mu,\delta\Pi_\nu\}=\delta^\mu_\nu$. For a given phase space function $f$, we consider its Taylor expansion around the background solution
\begin{equation*}
f(m,t)=\sum\limits_{k=0}^\infty f^{(k)}(\overline{m};\delta m,t),
\end{equation*}
where the function is allowed to have an explicit time dependence that will be necessary for the Dirac observables later. We introduced the abbreviations $m\coloneqq (q,p,N^\mu,\Pi_\mu)$, $\overline{m}\coloneqq(\overline{q},\overline{p},\overline{N}^\mu,\overline{\Pi}_\mu)$ and $\delta m\coloneqq(\delta q,\delta p, \delta N^\mu, \delta \Pi_\mu)$ denoting a generic point in the corresponding phase space, and $f^{(k)}$ denotes the $k$-th order of the Taylor expansion which is a polynomial of degree $k$ in the perturbations $(\delta q,\delta p, \delta N^\mu, \delta \Pi_\mu)$. Explicitly, we have
\begin{align}
f^{(k)}(\overline{m};\delta m,t) &= \!\!\!\!\!
\sum\limits_{\substack{\ell_1,\ldots,\ell_{10}=0 \\ \ell_1+\cdots +\ell_{10}=k}}^k \frac{1}{\ell_1!\cdots\ell_{10}!}\delta q^{\ell_1}\delta p^{\ell_2}\cdots \delta(\Pi_3)^{\ell_{10}}\left(\frac{\partial^{\ell_1+\cdots +\ell_{10}}f}{\partial \delta q^{\ell_1}\partial\delta p^{\ell_2}\cdots\partial\delta(\Pi_3)^{\ell_{10}}}\right)(\overline{m},t).\nonumber\\
\end{align}
The lemma E.1 from the appendix E of \cite{Giesel:2007wi} can be directly applied to our case, thus no generalization is needed here. For lemma E.2 in \cite{Giesel:2007wi}, we formulate the following generalization:
\begin{lemma}{(Generalization of lemma E.2 from \cite{Giesel:2007wi})}
\\
Suppose that ${\cal O}_f$ is an exact weak Dirac observable of the non-linear system under consideration with elementary variables $(q,p,N^\mu,\Pi_\mu)$, $\mu=0,\ldots 3$, with first class constraints $C_\mu(q,p)$, $\Pi_\mu$ as well as a canonical Hamiltonian $H_{\rm can}=N^\mu C_\mu(q,p)+\lambda^\mu\Pi_\mu$, where $\lambda^\mu$ are Lagrange multipliers. If we expand the canonical Hamiltonian up to $n$-th order in $(\delta q,\delta p, \delta N^\mu, \delta \Pi_\mu)$, with $n\geq 1$, the the following holds:
\begin{itemize} 
\item[{(i)}] the equations of motion to order $n$ of $\delta q, \delta p, \delta N^\mu, \delta \Pi_\mu$ are generated by
\begin{equation*}
H_{n}^{\rm can}\coloneqq\sum\limits_{k=2}^{n+1}H_{\rm can}^{(k)}.    
\end{equation*}
\item[{(ii)}]
the perturbation up to order $n$ of ${\cal O}_f$, given by
\begin{equation*}
{\cal O}_{f,n}\coloneqq\sum\limits_{k=0}^n {\cal O}^{(k)}_f,   
\end{equation*}
is a Dirac observable with respect to $H_{\rm can}$, and hence $C_\mu,\Pi_\mu$, up to terms of order at least $n+1$.
\end{itemize}
\end{lemma}
The proof of $(i)$ proceeds exactly with the same techniques as in \cite{Giesel:2007wi}, so we will just demonstrate this for the first variable. The remaining ones follow in a straightforward way. Let $m(t)=(q(t),p(t),N^\mu(t),\Pi_\mu(t))$ be an exact solution of the equations of motion, then we have
\begin{equation*}
\frac{\d q}{\d t}(t)=\left[\{q,H_{\rm can}\}\right]_{m=m(t)}.    
\end{equation*}
Subtracting the same equation for the background solution $\overline{m}(t)=(\overline{q}(t),\overline{p}(t),\overline{N}^\mu(t),\overline{\Pi}(t))$, we obtain
\begin{equation*}
\frac{\d\delta q}{\d t}(t)= \left[\{q,H_{\rm can}\}\right]_{m=m(t)} -\left[\{q,H_{\rm can}\}\right]_{m=\overline{m}(t)}  
=\sum\limits_{k=2}^\infty\frac{\partial H^{(k)}_{\rm can}}{\partial \delta p}    ~.
\end{equation*}
If we restrict to perturbations of order $n$, the statement $(i)$ follows directly. Likewise, this can be shown for all other phase space variables.

In order to prove $(ii)$, we consider that ${\cal O}^{(k)}_f$ depends explicitly on the background variables as well as time and we get
\begin{eqnarray}
\label{eq:TimDer1}
\frac{\d{\cal O}_f^{(k)}}{\d t}&=&
\frac{\partial{\cal O}_f^{(k)}}{\partial\overline{q}}\dot{\overline{q}}+\frac{\partial{\cal O}_f^{(k)}}{\partial \overline{p}}\dot{\overline{p}}+\frac{\partial{\cal O}_f^{(k)}}{\partial \overline{N}^\mu}\dot{\overline{N}}^\mu+\frac{\partial{\cal O}_f^{(k)}}{\partial \overline{\Pi}_\mu}\dot{\overline{\Pi}}_\mu\nonumber\\ 
&& 
+\frac{\partial{\cal O}_f^{(k)}}{\partial \delta{q}}\delta\dot{{q}}+\frac{\partial{\cal O}_f^{(k)}}{\partial \delta{p}}\delta\dot{{p}}+\frac{\partial{\cal O}_f^{(k)}}{\partial \delta{N}^\mu}\delta\dot{{N}}^\mu+\frac{\partial{\cal O}_f^{(k)}}{\partial \delta{\Pi}_\mu}\delta\dot{{\Pi}}_\mu+\frac{\partial {\cal O}_f}{\partial t},
\end{eqnarray}
where the `dot' refers to the evolution with respect to $H_{\rm can}$. Using the definition of the physical Hamiltonian, we can replace the time derivatives of the phase phase variables and rewrite (\ref{eq:TimDer1}) as,
\begin{eqnarray}
\label{eq:TimDer2}
\frac{\d{\cal O}_f^{(k)}}{\d t}&=&
\frac{\partial{\cal O}_f^{(k)}}{\partial\overline{q}}\frac{\partial H^{(1)}_{\rm can}}{\partial\delta p}-\frac{\partial{\cal O}_f^{(k)}}{\partial \overline{p}}\frac{\partial H^{(1)}_{\rm can}}{\partial\delta q}+\frac{\partial{\cal O}_f^{(k)}}{\partial \overline{N}^\mu}\frac{\partial H^{(1)}_{\rm can}}{\partial\delta \Pi_\mu}-\frac{\partial{\cal O}_f^{(k)}}{\partial \overline{\Pi}_\mu}\frac{\partial H^{(1)}_{\rm can}}{\partial\delta N^\mu}\\ 
&& 
+\frac{\partial{\cal O}_f^{(k)}}{\partial \delta{q}}\frac{\partial H^{\rm can}_n}{\partial\delta p}
-\frac{\partial{\cal O}_f^{(k)}}{\partial \delta{p}}\frac{\partial H^{\rm can}_n}{\partial\delta q}
+\frac{\partial{\cal O}_f^{(k)}}{\partial \delta{N}^\mu}\frac{\partial H^{\rm can}_n}{\partial\delta \Pi_\mu}
-\frac{\partial{\cal O}_f^{(k)}}{\partial \delta{\Pi}_\mu}\frac{\partial H^{\rm can}_n}{\partial\delta N^\mu}
+\frac{\partial {\cal O}_f}{\partial t},\nonumber
\end{eqnarray}
where we used that $\frac{\partial H_{\rm can}}{\partial q}\Big|_{m=\overline{m}}=\frac{\partial H^{(1)}_{\rm can}}{\partial \delta q}$ and likewise for the remaining variables. As the next step, we take a closer look at the partial derivatives with respect to the background degrees of freedom. We have
\begin{eqnarray*}
\frac{\partial{\cal O}^{(k)}_f}{\partial \overline{q}}
&=&
\sum\limits_{\substack{\ell_1,\cdots,\ell_{10}=0 \\ \ell_1+\cdots +\ell_{10}=k}}^k \frac{1}{\ell_1!\cdots\ell_{10}!}\delta q^{\ell_1}\delta p^{\ell_2}\cdots \delta(\Pi_3)^{\ell_{10}}\left(\frac{\partial^{\ell_1+\cdots +\ell_{10}+1}{\cal O}^{(k)}_f}{\partial \delta q^{\ell_1+1}\partial\delta p^{\ell_2}\cdots\partial\delta(\Pi_3)^{\ell_{10}}}\right)(\overline{m},t)\\
&=&
\frac{\partial}{\partial\delta q}\left(\sum\limits_{\substack{\ell_1,\cdots,\ell_{10}=0 \\ \ell_1+\cdots +\ell_{10}=k}}^k \frac{1}{(\ell_1+1)!\cdots\ell_{10}!}\delta q^{\ell_1+1}\delta p^{\ell_2}\cdots \delta(\Pi_3)^{\ell_{10}}\right. \\
&&
\left.\vphantom{\sum\limits_{\substack{\ell_1,\cdots,\ell_{10}=0 \\ \ell_1+\cdots +\ell_{10}=k}}^k}
\hphantom{\frac{\partial}{\partial\delta q}\Big(\sum\limits_{\substack{\ell_1,\cdots,\ell_{10}=0 \\ \ell_1+\cdots +\ell_{10}=k}}^k}
\cdot\left(\frac{\partial^{\ell_1+\cdots +\ell_{10}+1}{\cal O}^{(k)}_f}{\partial \delta q^{\ell_1+1}\partial\delta p^{\ell_2}\cdots\partial\delta(\Pi_3)^{\ell_{10}}}\right)(\overline{m},t)\right)
\\
&=&
\frac{\partial}{\partial\delta q}\left(\sum\limits_{\ell_1=1}^{k+1}\sum\limits_{\substack{\ell_2,\cdots,\ell_{10}=0 \\ \ell_1+\cdots +\ell_{10}=k+1}}^k \frac{1}{\ell_1!\cdots\ell_{10}!}\delta q^{\ell_1}\delta p^{\ell_2}\cdots \delta(\Pi_3)^{\ell_{10}} \right.\\
&&
\left.\vphantom{\sum\limits_{\substack{\ell_2,\cdots,\ell_{10}=0 \\ \ell_1+\cdots +\ell_{10}=k+1}}^k}
\hphantom{\frac{\partial}{\partial\delta q}\big(\sum\limits_{\ell_1=1}^{k+1}\sum\limits_{\substack{\ell_2,\cdots,\ell_{10}=0 \\ \ell_1+\cdots +\ell_{10}=k+1}}^k }
\cdot\left(\frac{\partial^{\ell_1+\cdots +\ell_{10}+1}{\cal O}^{(k)}_f}{\partial \delta q^{\ell_1}\partial\delta p^{\ell_2}\cdots\partial\delta(\Pi_3)^{\ell_{10}}}\right)(\overline{m},t)\right)
\\
&=&
\frac{\partial}{\partial\delta q}\left(
{\cal O}_f^{(k+1)}-\frac{1}{(k+1)!}\left(\frac{\partial^{k+1}O^{(k)}_f}{\partial\delta p^{k+1}}\delta p^{k+1}+\frac{\partial^{k+1}O^{(k)}_f}{\partial\delta (N^\mu)^{k+1}}\delta( N^\mu)^{k+1} \right.\right.\\
&&
\hphantom{\frac{\partial}{\partial\delta q}\big(
{\cal O}_f^{(k+1)}-\frac{1}{(k+1)!}\Big(}
\left.\left.\quad +\frac{\partial^{k+1}O^{(k)}_f}{\partial\delta (\Pi_\mu)^{k+1}}\delta( \Pi_\mu)^{k+1}
\right)\right)
\\
&=&
\frac{\partial O_{f}^{(k+1)}}{\partial\delta q},
\end{eqnarray*}
where we used in the last line that the second term in the bracket in the line before does not depend on $\delta q$. We can repeat the same calculation for the remaining variables to obtain
\begin{equation*}
\frac{\partial{\cal O}^{(k)}_f}{\partial \overline{p}}=\frac{\partial O_{f}^{(k+1)}}{\partial\delta p},\quad
\frac{\partial{\cal O}^{(k)}_f}{\partial \overline{N}^\mu}=\frac{\partial O_{f}^{(k+1)}}{\partial\delta N^\mu},\quad
\frac{\partial{\cal O}^{(k)}_f}{\partial \overline{\Pi}_\mu}=\frac{\partial O_{f}^{(k+1)}}{\partial\delta \Pi_\mu}.
\end{equation*}
Reinserting this back into the first line of the right hand side of (\ref{eq:TimDer2}), we realize that the first line precisely combines to the Poisson bracket $\{{\cal O}_f^{(k+1)},H^{(1)}_{\rm can}\}$ on the extended phase space. Considering this together with our former results, this leads to
\begin{equation*}
\frac{\d{\cal O}^{(k)}_f}{\d t}=\{{\cal O}_f^{(k+1)},H^{(1)}_{\rm can}\}+\{O^{(k)}_f,H_n^{\rm can}\}+\frac{\partial{\cal O}^{(k)}_f}{\partial t}   
\end{equation*}
which finally yields
\begin{equation}
\label{eq:TimeDer3}
\frac{\d{\cal O}_{f,n}}{\d t}=\sum\limits_{k=1}^n\left(\{{\cal O}_f^{(k+1)},H^{(1)}_{\rm can}\}+\sum\limits_{\ell=2}^{n+1}\{O^{(k)}_f,H^{(\ell)}_{\rm can}\}\right)
+\frac{\partial{\cal O}_{f,n}}{\partial t}.   
\end{equation}
Our aim is now to show that the R.H.S. of (\ref{eq:TimeDer3}) is up to terms of order $\delta^{n+1}$ given by the partial derivative with respect to $t$ only. Following \cite{Giesel:2007wi}, we realize that $\{O_f^{(k)},H^{(\ell)}_{\rm can}\}$ is of order $k+\ell-2$ and thus for fixed $k$,
 we can restrict the range of the sum over $\ell$ to $[2,n+2-k]$ -- up to terms that are at least of order $\delta^{n+1}$. In a second step, we will change the summation variable to $r=\ell+k-2$. This yields
 \begin{eqnarray*}
 \frac{\d{\cal O}_{f,n}}{\d t}&=&
 \sum\limits_{k=1}^n\left(\{{\cal O}_f^{(k+1)},H^{(1)}_{\rm can}\}+\sum\limits_{\ell=2}^{n+2-k}\{O^{(k)}_f,H^{(\ell)}_{\rm can}\}\right)
+\frac{\partial{\cal O}_{f,n}}{\partial t}+O(\delta^{n+1}) \\
&=&
 \sum\limits_{k=1}^n\left(\{{\cal O}_f^{(k+1)},H^{(1)}_{\rm can}\}+\sum\limits_{r=k}^{n}\{O^{(k)}_f,H^{(r-k+2)}_{\rm can}\}\right)
+\frac{\partial{\cal O}_{f,n}}{\partial t}+O(\delta^{n+1})\\
&=&
 \sum\limits_{r=1}^n\left(\{{\cal O}_f^{(r+1)},H^{(1)}_{\rm can}\}+\sum\limits_{k=1}^{r}\{O^{(k)}_f,H^{(r-k+2)}_{\rm can}\}\right)
+\frac{\partial{\cal O}_{f,n}}{\partial t}+O(\delta^{n+1}) \\
&=&
\sum\limits_{r=1}^n\sum\limits_{k=1}^{r+1}\{{\cal O}_f^{(k)},H^{(r-k+2)}_{\rm can}\}
+\frac{\partial{\cal O}_{f,n}}{\partial t}+O(\delta^{n+1}).
 \end{eqnarray*}
 At this point we take into account the fact that ${\cal O}_f$ is a Dirac observable of the full non-linear theory, that means
 \begin{equation}
 \label{eq:CondObsNL}
0\approx \{{\cal O}_f,H_{\rm can}\}=\sum\limits_{k,\ell=0}^\infty \{{\cal O}^{(k)}_f,H^{(\ell)}_{\rm can}\}
=
\sum\limits_{r=0}^\infty\sum\limits_{k=1}^{r+1}\{{\cal O}_f^{(k)},H^{(r-k+2)}_{\rm can}\}.
 \end{equation}
 Now, we can apply the same argument as in \cite{Giesel:2007wi}. Since the last equation is a weak identity on the entire phase space, the coefficients in the Taylor expansion of $\delta q ^\ell_1\cdots \delta(\Pi_3)^{\ell_{10}}$ have to vanish separately for all $\ell_1,\ldots,\ell_{10}\geq 0$. The term corresponding to order $r$ in (\ref{eq:CondObsNL}) contains all terms of the kind $\delta q ^{\ell_1}\cdots\delta\Pi_3^{\ell_{10}}\Big|_{\ell_1+\cdots +\ell_{10}=r}$. Consequently, we can can conclude that for all $r$ we have
 \begin{equation*}
\sum\limits_{k=1}^{r+1} \{{\cal O}^{(k)}_f,H_{\rm can}^{(r-k+2)}\}\approx 0     
 \end{equation*}
 and this further implies
\begin{equation}
\label{eq:TimeDerFinal}
  \frac{d{\cal O}_{f,n}}{dt}\approx \frac{\partial{\cal O}_{f,n}}{\partial t}  +O(\delta^{n+1}).
\end{equation} 
Hence, we have shown that $(ii)$ in {{lemma 4}} holds true. That is,  a Dirac observable in perturbation theory up to order $n$ is still a Dirac observable up to corrections of order $\delta^{n+1}$.

Let us finally consider the case $n=1$ that is relevant for the linear perturbation theory. Then, the corrections $O(\delta^{n+1})$ vanish identically, which happens only for $n=1$. In this case, we obtain
\begin{equation*}
 \frac{\d{\cal O}_{f,n}}{\d t}\approx \frac{\partial{\cal O}_{f,n}}{\partial t}    
\end{equation*}
and the non-linear condition $\{{\cal O}_f,H_{\rm can}\}\approx 0$ carries over to 
\begin{eqnarray}
\label{eq:ConObsFinal}
a.) && \{O^{(1)}_f,H^{(1)}_{\rm can}\}\approx 0 \\
b.) && \{O^{(2)}_f,H^{(1)}_{\rm can}\}+\{O^{(1)}_f,H^{(2)}_{\rm can}\}\approx 0,
\end{eqnarray}
where we also included the condition a.) that is required for linearized observables on the linearized phase space. Furthermore, this result shows that observables only have a non-trivial evolution if they are explicitly time-dependent. Whether this evolution can again be written in Hamiltonian form by means of a physical Hamiltonian is an additional question that has been positively answered in the deparametrized matter models for instance in \cite{Giesel:2007wi,Giesel:2007wk,Domagala:2010bm,Husain:2011tm,Giesel:2012rb,Giesel4,Giesel:2017mfc}. In the context of the relational formalism, one can then reexpress the evolution of the observables ${\cal O}_f$ in terms of the physical time parameter $\tau$ by using appropriate Jacobians.

This lemma further shows that even for linearized perturbation theory we need to know ${\cal O}^{(2)}_f$ in order to actually test condition b.). Also, a question arises on why it was justified to not consider such contributions in the derivation of the equations of motion of the gauge variant quantities in the extended ADM phase space because even for non-observables the equation (\ref{eq:TimeDer3}) still holds. The reason is simple because we only computed equations of motion for elementary phase space variables for which ${\cal O}^{(2)}_f$ trivially vanishes. However, this is no longer given if we, for instance, consider more complicated observables such as the Bardeen potential or the Mukhanov-Sasaki variable that will have a non-trivial contribution in second order perturbation theory. Note that this is also not relevant for our results in this work since we used the lemma 4 derived in this article to compute the equations of motion for such observables in a different way using the equations of motion of the gauge variant phase space variables.


\begin{thebibliography}{10}

\bibitem{Pons3}
J.~M. Pons, D.~C. Salisbury, and K.~A. Sundermeyer,
\newblock Observables in classical canonical gravity: folklore demystified,
\newblock {\em J. Phys.: Conf. Ser.}, 222:012018, 2010,
\newblock \href{https://arxiv.org/abs/1001.2726v2}{arXiv:1001.2726v2}.

\bibitem{Pons4}
J.~M. Pons, D.~C. Salisbury, and K.~A. Sundermeyer,
\newblock Revisiting observables in generally covariant theories in the light
  of gauge fixing methods,
\newblock {\em Phys. Rev. D}, 80:084015, 2009,
\newblock \href{https://arxiv.org/abs/0905.4564v2}{arXiv:0905.4564v2}.

\bibitem{bardeen}
J.~M. Bardeen,
\newblock {Cosmological perturbations from quantum fluctuations to large scale
  structure},
\newblock In {\em {Cosmology and particle physics. Proceedings, CCAST (World
  Laboratory) Symposium/Workshop, Nanjing, P.R. China, June 30 - July 12,
  1988}}, pages 1--64, 1988.

\bibitem{kodama-sasaki}
H. Kodama and M. Sasaki,
\newblock {Cosmological Perturbation Theory},
\newblock {\em Prog. Theor. Phys. Suppl.}, 78:1--166, 1984.

\bibitem{Mukhanov}
V.~F. Mukhanov, H.~A. Feldman, and R.~H. Brandenberger,
\newblock Theory of cosmological perturbations,
\newblock {\em Phys. Rep.}, 215(5-6):203--333, 1992.

\bibitem{Ellis:1989jt}
G.~F.~R. Ellis and M.~Bruni,
\newblock {Covariant and Gauge Invariant Approach to Cosmological Density
  Fluctuations},
\newblock {\em Phys. Rev.}, D40:1804--1818, 1989.

\bibitem{Langlois}
D.~Langlois,
\newblock {Hamiltonian formalism and gauge invariance for linear perturbations
  in inflation},
\newblock {\em Class. Quant. Grav.}, 11:389--407, 1994.

\bibitem{Anderegg:1994xq}
S.~Anderegg and V.~F. Mukhanov,
\newblock {Path integral quantization of cosmological perturbations},
\newblock {\em Phys. Lett.}, B331:30--38, 1994.

\bibitem{Giesel:2017roz}
K. Giesel and A. Herzog,
\newblock {Gauge invariant canonical cosmological perturbation theory with
  geometrical clocks in extended phase-space — A review and applications},
\newblock {\em Int. J. Mod. Phys.}, D27(08):1830005, 2018.

\bibitem{Giesel:2018opa}
K. Giesel, A. Herzog, and P. Singh,
\newblock {Gauge invariant variables for cosmological perturbation theory using
  geometrical clocks},
\newblock {\em Class. Quant. Grav.}, 35(15):155012, 2018.

\bibitem{Malkiewicz:2018ohk}
P. Ma\l{}kiewicz,
\newblock {Hamiltonian formalism and gauge-fixing conditions for cosmological
  perturbation theory}, 	arXiv:1810.11621 [gr-qc]

\bibitem{RovelliPartial}
C.~Rovelli,
\newblock Partial observables,
\newblock {\em Phys. Rev. D}, 65:124013, 2002,
\newblock \href{https://arxiv.org/abs/gr-qc/0110035}{gr-qc/0110035}.

\bibitem{RovelliObservable}
C.~Rovelli,
\newblock What is observable in classical and quantum gravity?,
\newblock {\em Class. Quant. Grav.}, 8(2):297, 1991.

\bibitem{Dittrich}
B.~Dittrich,
\newblock Partial and complete observables for hamiltonian constrained systems,
\newblock {\em Gen. Rel. Grav.}, 39:1891--1927, 2004,
\newblock \href{https://arxiv.org/abs/gr-qc/0411013}{gr-qc/0411013}.

\bibitem{Dittrich2}
B.~Dittrich,
\newblock Partial and complete observables for canonical general relativity,
\newblock {\em Class. Quant. Grav.}, 23:6155--6184, 2005,
\newblock \href{https://arxiv.org/abs/gr-qc/0507106v1}{gr-qc/0507106v1}.

\bibitem{Giesel:2007wi}
K.~Giesel, S.~Hofmann, T.~Thiemann, and O.~Winkler,
\newblock Manifestly gauge-invariant general relativistic perturbation theory:
  I. foundations,
\newblock {\em Class. Quant. Grav.}, 27:055005, 2010,
\newblock \href{https://arxiv.org/abs/0711.0115v2}{arXiv:0711.0115v2}.

\bibitem{Giesel:2007wk}
K.~Giesel, S.~Hofmann, T.~Thiemann, and O.~Winkler,
\newblock Manifestly gauge-invariant general relativistic perturbation theory:
  Ii. frw background and first order,
\newblock {\em Class. Quant. Grav.}, 27:055006, 2010,
\newblock \href{https://arxiv.org/abs/0711.0117v1}{arXiv:0711.0117v1}.

\bibitem{Han:2015jsa}
Y.~Han, K. Giesel, and Y. Ma,
\newblock {Manifestly gauge invariant perturbations of scalar–tensor theories
  of gravity},
\newblock {\em Class. Quant. Grav.}, 32:135006, 2015.

\bibitem{Giesel:2009jp}
K. Giesel, J. Tambornino, and T. Thiemann,
\newblock {LTB spacetimes in terms of Dirac observables},
\newblock {\em Class. Quant. Grav.}, 27:105013, 2010.

\bibitem{Dittrich-Tambornino2}
B. Dittrich and J. Tambornino,
\newblock {Gauge invariant perturbations around symmetry reduced sectors of
  general relativity: Applications to cosmology},
\newblock {\em Class. Quant. Grav.}, 24:4543--4586, 2007.

\bibitem{Giesel:2007wn}
K.~Giesel and T.~Thiemann,
\newblock {Algebraic quantum gravity (AQG). IV. Reduced phase space
  quantisation of loop quantum gravity},
\newblock {\em Class. Quant. Grav.}, 27:175009, 2010.

\bibitem{Domagala:2010bm}
M.~Domagala, K.~Giesel, W.~Kaminski, and J.~Lewandowski,
\newblock Gravity quantized: Loop quantum gravity with a scalar field,
\newblock {\em Phys. Rev. D}, 22:104038, 2010,
\newblock \href{https://arxiv.org/abs/1009.2445}{arXiv:1009.2445}.

\bibitem{Pons1}
J.~M. Pons and J.~Antonio Garcia,
\newblock Rigid and gauge noether symmetries for constrained systems,
\newblock {\em Int. J. Mod. Phys. A}, 15:4681--4721, 2000,
\newblock \href{https://arxiv.org/abs/hep-th/9908151v2}{hep-th/9908151v2}.

\bibitem{Pons2}
J.M. Pons, D.C. Salisbury, and L.C. Shepley,
\newblock Gauge transformations in the lagrangian and hamiltonian formalisms of
  generally covariant theories,
\newblock {\em Phys. Rev. D}, 55(2):658, 1996,
\newblock \href{https://arxiv.org/abs/gr-qc/9612037v1}{gr-qc/9612037v1}.

\bibitem{Giesel:2012rb}
K. Giesel and T. Thiemann,
\newblock {Scalar Material Reference Systems and Loop Quantum Gravity},
\newblock {\em Class. Quant. Grav.}, 32:135015, 2015.

\bibitem{Giesel4}
K.~Giesel and A.~Oelmann,
\newblock Reduced loop quantization with four klein-gordon scalar fields as
  reference matter,
\newblock 2016,
\newblock \href{https://arxiv.org/abs/1610.07422v1}{arXiv:1610.07422v2}, submitted to CQG.

\bibitem{Thiemann2}
T.~Thiemann,
\newblock Reduced phase space quantization and dirac observables,
\newblock {\em Class. Quant. Grav.}, 23:1163--1180, 2006,
\newblock \href{https://arxiv.org/abs/gr-qc/0411031v1}{gr-qc/0411031v1}.

\bibitem{Brandenberger}
R.~H. Brandenberger,
\newblock {Lectures on the theory of cosmological perturbations},
\newblock {\em Lect. Notes Phys.}, 646:127--167, 2004.

\bibitem{Husain:2011tm}
V. Husain and T. Pawlowski,
\newblock {Dust reference frame in quantum cosmology},
\newblock {\em Class. Quant. Grav.}, 28:225014, 2011.

\bibitem{Giesel:2017mfc}
  K.~Giesel and A.~Oelmann,
  \newblock {Comparison Between Dirac and Reduced Quantization in LQG-Models with Klein-Gordon Scalar Fields},
  \newblock {\em Acta Phys.\ Polon.\ Supp.},10:38, 2017.


\end{thebibliography}

\end{document}